\documentclass[5p,times]{elsarticle}

\usepackage{wrapfig}
\usepackage{pdfpages}
\usepackage{float}
\usepackage{parskip}
\usepackage{xcolor}
\usepackage{chngcntr}
\usepackage{array}
\usepackage{url}
\usepackage{hyperref}
\usepackage{todonotes}
\usepackage{booktabs}
\usepackage{picins}
\counterwithin{figure}{section}
\counterwithin{table}{section}
\begin{document}
\begin{frontmatter}

\title{Serverless Containers -- rising viable approach to Scientific Workflows}

\author[1]{Krzysztof Burkat}
\ead{burkat.krzysztof@gmail.com}
\author[1]{Maciej Pawlik}
\ead{maciej.pawlik@cyfronet.pl}
\author[1]{Bartosz Balis}
\ead{balis@agh.edu.pl}
\author[1]{Maciej Malawski}
\ead{malawski@agh.edu.pl}
\author[2]{Karan Vahi}
\ead{vahi@isi.edu}
\author[2]{Mats Rynge}
\ead{rynge@isi.edu}
\author[2]{Rafael Ferreira da Silva}
\ead{rafsilva@isi.edu}
\author[2]{Ewa Deelman}
\ead{deelman@isi.edu}

\address[1]{AGH University of Science and Technology, Department of Computer Science, Krakow, Poland}
\address[2]{University of Southern California, Information Sciences Institute, Marina del Rey, CA, USA}

\author{
    % \IEEEauthorblockN{
    %     Krzysztof Burkat\IEEEauthorrefmark{1}, 
    %     Maciej Pawlik\IEEEauthorrefmark{1}, 
    %     Bartosz Balis\IEEEauthorrefmark{1}, 
    %     Maciej Malawski\IEEEauthorrefmark{1},
    %     Karan Vahi\IEEEauthorrefmark{3}, 
    %     Mats Rynge\IEEEauthorrefmark{3},\\
    %     Rafael Ferreira da Silva\IEEEauthorrefmark{3}, 
    %     Ewa Deelman\IEEEauthorrefmark{3}
    %     \\
    % }
    % \IEEEauthorblockA{
    % \IEEEauthorrefmark{1}AGH University of Science and Technology, Department of Computer Science, Krakow, Poland \\
    % \IEEEauthorrefmark{3}University of Southern California, Information Sciences Institute, Marina del Rey, CA, USA \\
    % Emails: \{malawski,balis\}@agh.edu.pl, \{vahi,rynge,rafsilva,deelman\}@isi.edu
    % }
}

% \maketitle

\begin{abstract}
Increasing popularity of the serverless computing approach has led to the emergence of new cloud infrastructures working in Container-as-a-Service (CaaS) model like AWS Fargate, Google Cloud Run, or Azure Container Instances. They introduce  an innovative approach to running cloud containers where developers are freed from managing underlying resources. In this paper, we focus on evaluating capabilities of elastic containers and their usefulness for scientific computing in the scientific workflow paradigm using AWS Fargate and Google Cloud Run infrastructures. For experimental evaluation of our approach, we extended HyperFlow engine to support these CaaS platform, together with adapting four real-world scientific workflows composed of several dozen to over a hundred of tasks organized into a dependency graph. We used these workflows to create cost-performance benchmarks and flow execution plots, measuring delays, elasticity, and scalability. The experiments proved that serverless containers can be successfully applied for scientific workflows. Also, the results allow us to gain insights on specific advantages and limits of such platforms.
\end{abstract}
\end{frontmatter}
\section{Introduction}
Serverless computing is a general execution model for distributed applications running on the cloud~\cite{Castro2019,baldini2017serverless}. In this model, the responsibility for managing underlying resources rests solely on the cloud provider, therefore it allows developers to focus on the user side of development, simplifying the process of creating, deploying, and running applications. The serverless architecture supports elasticity and scalability, meaning it can provision and de-provision many cloud instances dynamically on the fly, depending on the demand. These features of serverless computing make it an attractive execution model for scientific workflows. Such workflows usually consist of multiple stages, each of  which may contain from several to even thousands of tasks, thus elasticity is a~key requirement. The last important serverless feature is its fine-grained pay-as-you-go pricing model wherein users are only charged for what they actually use with typical accuracy of 100ms. Usually, the cost is calculated as the amount of used resources (e.g., memory) multiplied by the amount of execution time units and by the price per execution time unit. This approach frees developers from manual management of resource allocation and can save money in comparison to on-premises infrastructure.

A prime example of serverless approach is Function-as-a-Service (FaaS) model. In this model, developers write single-purpose functions in various programming languages, which are then deployed onto cloud platforms and can be triggered by specified events to perform a specified action. It is therefore an event-based architecture. FaaS is sometimes refereed as serverless because of its simplicity and elasticity. Most popular FaaS implementations are AWS Lambda\footnote{https://aws.amazon.com/lambda/}, Google Cloud Functions\footnote{https://cloud.google.com/functions/}, Azure Functions\footnote{https://azure.microsoft.com/en-us/services/functions/}, and IBM Cloud Functions\footnote{https://cloud.ibm.com/functions/}. All of them can execute binary code, which makes them a viable choice for general purpose computing tasks that form a scientific workflow.

Another, relatively new in cloud computing, model is Container-as-a-Service (CaaS). 
Its main feature is the utilization of a container-based virtualization. In comparison to FaaS, CaaS adds additional layer in a form of a container, which bundles user applications. The container is deployed onto the cloud platform and when triggered, container instances are provisioned and de-provisioned on the fly depending on current demands. Container wrapping adds additional complexity, but it also gives more capabilities for configuring runtime environment and executing applications. AWS Fargate, Google Cloud Run, and Azure Container Instances are the main examples of such elastic containers. Similarly to FaaS, CaaS can be utilized for scientific workflow executions, as we show here.

In this paper, our goal is to determine usefulness of a Container-as-a-Service model  for executing scientific workflows on a CaaS, such as the Fargate or Cloud Run services. We also aim to characterize their capabilities and possible challenges. We present an experimental evaluation using the HyperFlow workflow engine~\cite{hflow-fgcs16} and four real-world scientific workflows that execute on these infrastructures.

The main contributions of the paper are threefold:
\begin{itemize}
    \item We discuss the advantages and limitations of serverless containers and their application to scientific workflows.
    \item We extend the HyperFlow workflow engine to support Fargate and Cloud Run services.
    \item We evaluate the proposed approach using four real-life scientific workflow applications of various characteristics and measure the performance of the container platforms.
\end{itemize}
To our best knowledge this paper is the first attempt to use serverless containers for scientific workflows and also the first performance evaluation of the Fargate and Cloud Run platforms. Note that our main focus is not to simply compare these platforms as they may evolve, but rather show their general features and limitations when running scientific workflows. Additionally, we show how serverless containers may  complement other existing serverless services (e.g., Lambda) in a hybrid approach.

The structure of the paper is as follows: in Section~\ref{sec:related-work} we give an overview about recent developments regarding using serverless services for scientific workflows. Next, in Section~\ref{sec:serverless-clouds} we discuss in detail the challenges related to adapting scientific workflows to serverless platforms and various potential models of execution. Section~\ref{sec:containers} presents our approach to these challenges, which utilizes the HyperFlow engine with Fargate and Cloud Run services. In Section~\ref{sec:experiments}, we present the results of workflow execution along with cost-performance benchmarks. Section~\ref{sec:conclusion} concludes the paper, and identifies directions for future research.

\section{Related Work}
\label{sec:related-work}

In our earlier work, we conducted an evaluation of serverless cloud functions with scientific workflows~\cite{malawski-2017-hf-serverless} on AWS Lambda and Google Cloud Functions using HyperFlow, a lightweight workflow engine. Another more advanced solution was introduced in~\cite{Jiang2017} where the authors combined IaaS and FaaS models in their DEWE workflow engine to show the benefits of such a hybrid approach. 
In~\cite{benchmarking-1} and~\cite{benchmarking-2}, we presented a performance evaluation of major cloud function providers: AWS Lambda, Azure Functions, Google Cloud Functions, and IBM Cloud Functions by running various benchmarks on these infrastructures.
All the works referenced above were focused on cloud functions, while in this paper we address serverless containers, which provide different capabilities and require a different approach.

Much work had been done in the area of workflow scheduling in the cloud. One approach is Deadline-Budget Workflow Scheduling (DBWS)~\cite{static-scheduling}, a heuristic scheduling algorithm which takes into account two constraints -- time and cost. In~\cite{kijak-SDBWS}, we improved this algorithm for serverless infrastructures. Another serverless-based approach is presented in~\cite{mujezinovic-serverless-scheduling} where the authors leverage Fargate service and producer-consumer pattern to introduce a fully serverless and infinitely scalable architecture. An interesting discussion of resource allocation for multi-tenant serverless computing platforms is given in~\cite{kim-cpu-heuristic}, where the authors focus on workload fluctuations and performance degradation in these platforms and propose CPU cap control heuristic as a remedy.

There are many approaches to a way of executing workflows, especially considering various infrastructures. In~\cite{mesos-makeflow}, Mesos and Makeflow are utilized to connect workflow system to container-based schedulers. There are also more advanced execution environments. Examples include Skyport~\cite{skyport} -- container-based workflow orchestrator; and Endofday~\cite{endofday} --  a workflow engine that orchestrates a directed acyclic graph (DAG) of science applications on containers. A more general and scalable solution to workflow execution is provided by the Pegasus Workflow Management System~\cite{deelman-pegasus-1, deelman-pegasus-2}, which goal is to map abstract workflow descriptions to various distributed computing infrastructures. However, all these approaches rely on on-premise or cloud-based clusters, not on serverless container platforms as we do here.

Although serverless capabilities are still a relatively new paradigm in cloud world, there are many potential applications for such environments. A prime example comes from the video industry where the need for compute power is enormous~\cite{ExCamera}. These applications also need low-latency and massively parallel video processing systems utilizing cloud function services~\cite{Sprocket}. Another example comes from the seismic imaging use case~\cite{Witte2020}, where an event-based serverless architecture is presented. Other examples include on-premises serverless computing for event-driven data processing applications~\cite{on-premises-serverless} or a framework for the management of the Internet of Things devices  based on the serverless computing paradigm~\cite{serverless-iot}. All these approaches use FaaS, while we focus on CaaS model.

None of the aforementioned related work considers utilizing serverless containers for running scientific workflows and we have not found any other works with substantial research on this topic. In this paper, we address this interesting research matter.
\section{Scientific Workflows in Serverless Clouds}
\label{sec:serverless-clouds}

In this section, we present serverless clouds on the example of Lambda, Fargate and Cloud Run, and discuss their implications for resource management in scientific workflows.

\subsection{Serverless Clouds}

Serverless computing is an execution model, where the cloud provider is responsible for allocating server-side resources dynamically, which can be used by users. This model puts emphasis on the business side of development, simplifying the process of creating, deploying, and running applications by freeing programmers from having to maintain servers. In addition, the serverless architecture supports scalability and elasticity which means that developers do not need to manage auto-scaling policies.

In the FaaS model, developers are provided with a platform where a piece of code known as single-purpose function can be written, deployed, and later triggered by events. Usually the set of supported languages is limited by the execution environment over which users do not have control. At most users can use package managers to install custom libraries, but in some situations, this solution is not sufficient. When deployed, the cloud functions can be raised by an event from cloud infrastructure or direct HTTP request. 
The functions are very elastic, they can adapt to changing workload (number of events of requests) by provisioning and de-provisioning resources on the fly. The number of running function instances can vary from several to thousands. 
The pricing model usually depends on number of requests and compute time and differs between various cloud providers.

 In container-based virtualization, users are provided with a platform where the unit of execution, a container, can be managed and deployed. Such containers can be later triggered by a proper API, and when it happens, they are provisioned and de-provisioned at runtime depending on the changing workload, and underlying resources are managed automatically. In this container-based solution developers have full control over the execution environment in the container thus they can choose any operating system, libraries, or preferred programming language. They also define a starting point for the container. The price is usually calculated based on the CPU and memory resources used per unit of time.

\begin{table*}[!t]
\begin{center}
\caption{Comparison of chosen cloud services.}
\begin{tabular}{ >{\arraybackslash}m{3.7cm} >{\centering\arraybackslash}m{4.2cm} >{\centering\arraybackslash}m{4.2cm} >{\centering\arraybackslash}m{4.2cm}} 
 \toprule
 \textbf{} & \textbf{AWS Lambda} & \textbf{AWS Fargate} & \textbf{Google Cloud Run} \\ 
 \midrule
 Execution environment & Amazon Linux &  User defined &  User defined \\
 \midrule
 Supported languages & Java, Python, Node.js, Go, Ruby, C\# &  Depends on execution environment &  Depends on execution environment \\
 \midrule
 Memory allocation & From 128 MB to 3008 MB & From 0.5 GB to 30 GB  & From 128 MiB to 2 GiB\\
 \midrule
 CPU allocation & Automatic (AWS controlled) & From 0.25 to 4 virtual cores  & From 1 to 2 virtual cores\\
 \midrule
 Disk space  & 512 MB & 10 GB & Uses memory\\
 \midrule
 Maximum execution time  & 900s  & No limit & 900s\\
 \midrule
 Maximum parallel executions  & 1000 & 100 & 1000\\
 \midrule
 Deployment unit  & Zipped code & Container & Container\\
 \bottomrule
\end{tabular}
\label{lambda-fargate-gcloud-comparison}
\end{center}
\end{table*}

Table~\ref{lambda-fargate-gcloud-comparison} shows key differences between serverless services we used for our experiments. We can see that even though Cloud Run is based on containers, it has similar limits to Lambda which works in FaaS model. On the other hand, Fargate is more robust in terms of limits, but it comes at a price of reduced parallel executions as compared to Lambda and Cloud Run. From this table, we can conclude that the main CaaS advantage is the additional deployment (e.g., execution environment) capabilities.

\subsection{Implications for Resource Management}
\label{sec:implications}

As we can see from Table~\ref{lambda-fargate-gcloud-comparison}, using serverless containers (such as Fargate or Cloud Run) has some advantages over cloud functions (such as Lambda), but there are still important questions and decisions regarding resource management of workflows when using these infrastructures. We discuss these issues below.

\subsubsection*{How to map tasks to containers} 
Serverless container platforms such as Fargate or Cloud Run provide higher level of abstraction for their users than traditional clusters with container support (either on-premise, or cloud based), using Docker or Kubernetes. When running workflows on non-serverless clusters, the user or the workflow management system is responsible for provisioning physical or virtual machines on which the containers are executed. Finding the right size of such a cluster is non trivial for the workflows where the resource demands varies during execution, e.g. when a highly parallel stage with many tasks is followed by a reduction phase with a small number of tasks. Auto-scaling of such clusters is possible, either at the level of VMs or containers, but non trivial~\cite{OrzechowskiUCC18}. In contrast, serverless containers take care of automatic resource provisioning and auto-scaling of the underlying cluster, so the workflow system does not need to deal with these problems.

\begin{figure}[H]
\centering
\includegraphics[width=\columnwidth]{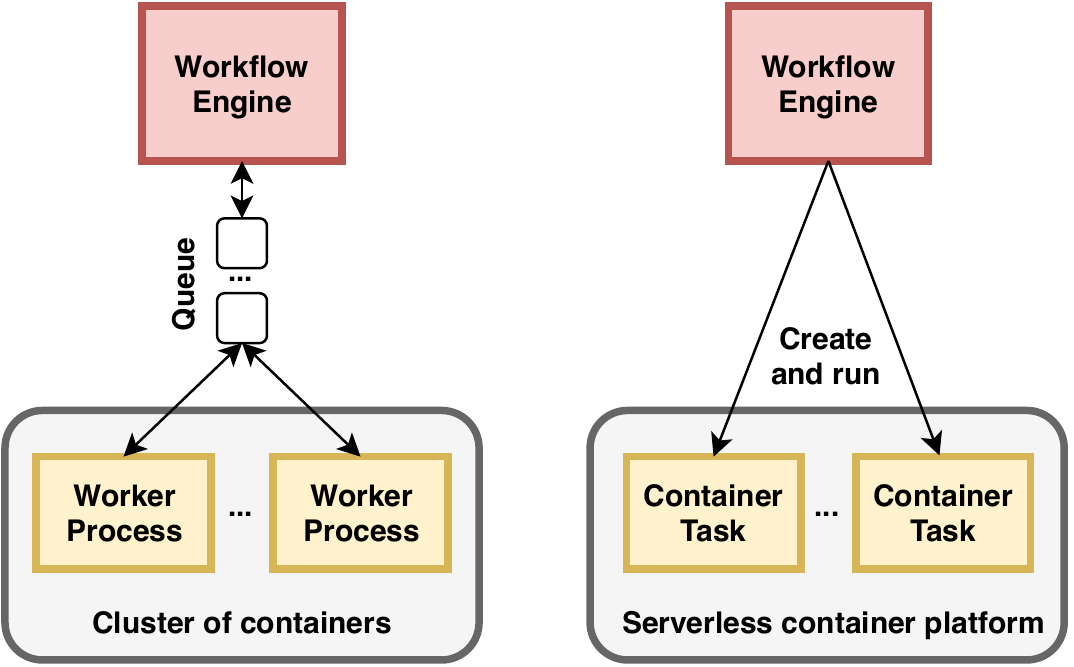}
\caption{Execution models of workflow tasks on containers}
\label{fig:models}
\end{figure}

There is still, however, a design decision on how to map the workflow tasks to containers. When running workflows on clusters or clouds, a workflow management system (WMS) typically runs a set of processes on the worker nodes, which communicate with the WMS via a queue and fetch the tasks ready for execution. For instance, this is the case for Pegasus, which uses Condor daemons to execute tasks. The same model is used by HyperFlow in IaaS cloud setup, where RabbitMQ is used as a queue and AMQP-Executor processes are running on the worker nodes, one per each VM in the cloud. In such model, one worker can execute multiple tasks during its lifetime, and the queue provides the load-balancing. This model is shown in Fig.~\ref{fig:models} on the left. Although this approach can be easily implemented for serverless containers, as the worker code can be simply reused, there are no benefits of the serverless computing model. If the container is a long-running service, we have to decide when to spawn and terminate it, so the auto-scaling decisions need to be made on the client-side to avoid under- or over-provisioning. On the contrary, the pure serverless model, as shown in Fig.~\ref{fig:models} on the right, does not require a queue, and assumes a \emph{one-to-one task to container mapping}, where for each task a separate container is spawned and then de-allocated after the task is complete. At first it seems naive, but such a mapping greatly simplifies the design (tasks isolation) and fully leverages the capabilities of the serverless computing model, where the cloud infrastructure is responsible for resource provisioning. We can say in other words that instead of implementing our own queue and autoscaling mechanisms, we rely on the autoscaling (provisioning) provided by the serverless infrastructure, and also on its internal queue which is used for buffering multiple requests transparently. Such a mapping  also mitigates resource contention within a container (too many tasks running in one container) and yields more deterministic execution times. On the other hand, it has additional cost of container startup overhead for each task, which we measure in Section~\ref{sec:experiments}, showing that it is  non-negligible, but not prohibitive for most of the use cases.

\subsubsection*{Which tasks to run on cloud functions and which on containers}

Despite many advantages of serverless containers over cloud functions, the latter model has clear advantages for some classes of workflow tasks. As cloud functions such as Lambda have a highly elastic provisioning model, i.e. they can spawn thousands of tasks almost immediately, they are well suited for short running tasks with high parallelism. Which size of a task is right for Lambda and which for Fargate or Cloud Run depends on the execution limits (as stated in  Tab.~\ref{lambda-fargate-gcloud-comparison}) and overheads of these specific infrastructures, and we measure them in our experiments described in Section~\ref{sec:experiments}.

\subsubsection*{How to allocate CPU resources to tasks} 

Fargate, Lambda, and Cloud Run, as other serverless platforms, provide control over CPU allocation to the computing tasks. In Lambda, the CPU share is proportional to the allocated RAM, while in Fargate and Cloud Run it can be adjusted within limits range (see Tab.~\ref{lambda-fargate-gcloud-comparison}). It means the important resource management decisions that influence time and cost of workflow execution are left for the user. As these decisions are non-trivial, they can be subject to scheduling research, and we have started the work on these problems, see~\cite{kijak-SDBWS}. Extending this scheduling work to serverless containers is beyond the scope of this paper, but the first step is to evaluate the performance of serverless containers in this platform, on which we report in Section~\ref{sec:experiments}.
\section{Experimental Framework}
\label{sec:containers}

Most of the existing scientific workflows are not designed to work in the elastic, container-based clouds. In order to run them there, they must be properly adapted beforehand. In this work, to evaluate our approach to workflows on serverless containers, we have developed such an adaptation for Fargate and Cloud Run services. Our solution is implemented as an extension to the HyperFlow engine.

\subsection{HyperFlow}

HyperFlow is a model of computation, programming approach, and enactment engine for scientific workflows~\cite{hflow-fgcs16}. It is the engine implemented in a scripting language -- JavaScript, which enables users to run scientific workflows in distributed computing infrastructures. The model of computation utilized by HyperFlow is based on Process Network family which has simple and concise syntax. Moreover, workflows are described as graphs in the JSON format with nodes representing workflow activities (jobs) and edges representing exchanged data.

\begin{figure}[H]
    \centering
    \includegraphics[width=\columnwidth]{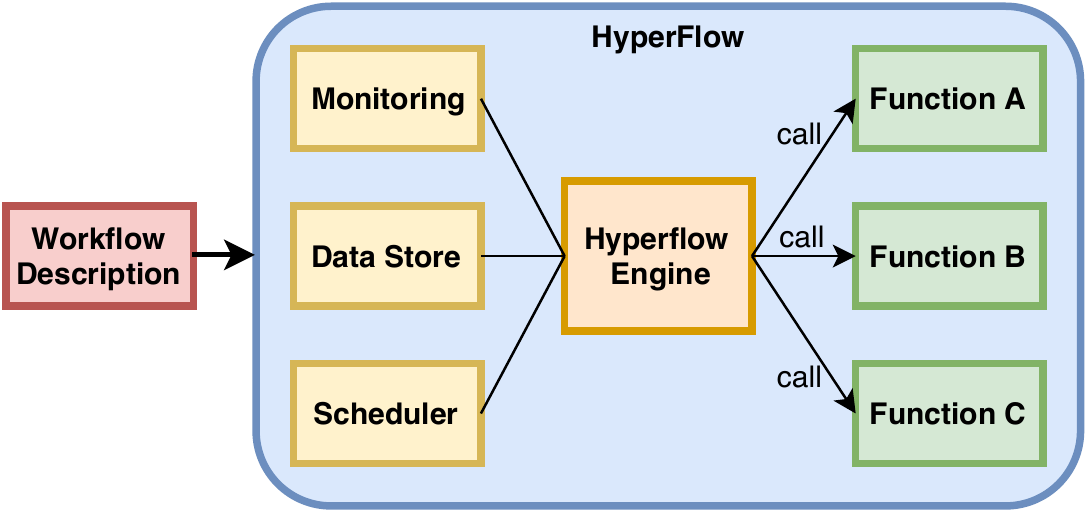}
    \caption{Architecture diagram of HyperFlow engine, where each task in a workflow is handled by calling a function, which can process it internally or delegate execution to an external service.}
    \label{HyperFlow}
\end{figure}

Thanks to the simplicity and expressiveness of workflow description, users can define and run complex workflow patterns. When HyperFlow parses a workflow, it recognizes all activities with their corespondent arguments to execute. It also supervises the order of execution and waits until all dependencies of a given activity are met. After that, it calls the Function defined by the user (or predefined one), which is an entry point for executing the actual scientific task related to a given workflow activity. This gives the flexibility as to how and when a given activity should be executed. The entry point functions can orchestrate execution of tasks on various distributed resources as shown in Figure~\ref{HyperFlow}. Each function can handle different kinds of execution. Thanks to that, various cloud computation models like FaaS or CaaS can be successfully applied. 

\subsection{Preparing the Workflow}

Before running any workflow, we need to have its representation in some format like JSON or XML. We chose the JSON format because it is supported by HyperFlow and because it has an easy to understand structure (name-value pairs).

To generate the workflows, we developed a set of scripts that have a recipe for the entire workflow -- it consists of all the dependencies between workflow's tasks. Besides that, the scripts take various arguments that determine the final structure of the workflow (e.g., number of tasks).

For some scientific workflows there are already dedicated generators. This was the case for some of the workflows included in our test suite. The problem was that those workflow descriptions used the XML-based DAG format from Pegasus (DAX). Fortunately, HyperFlow includes converters which allowed us, with small corrections, to transform XML documents to JSON format.

\subsection{Extending HyperFlow for Serverless Containers}

\begin{figure}[H]
    \centering
    \includegraphics[width=\columnwidth]{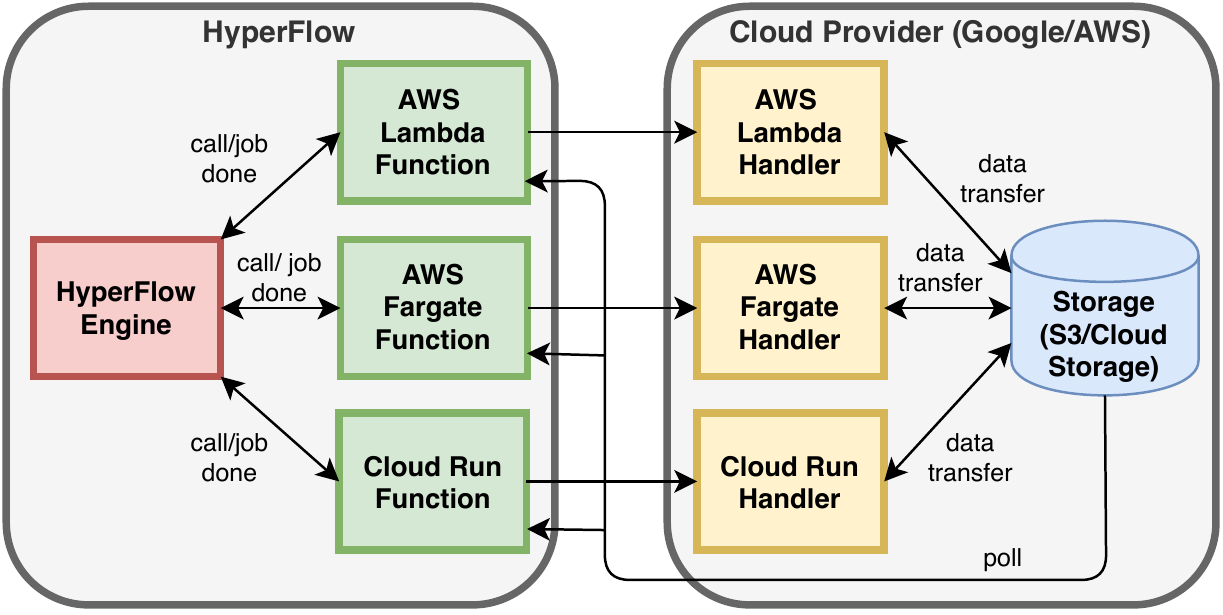}
    \caption{Proposed solution framework.}
    \label{framework}
\end{figure}

The HyperFlow engine provides support for executing workflows on real cloud services, like Lambda. It also serves as a buffer if many tasks are requested and provides retry mechanism (for each individual task). In order to execute tasks on Fargate or Cloud Run, we have extended the engine with additional capabilities to support the aforementioned systems. In Figure~\ref{framework}, we present the proposed solution framework, which has four key components:

\begin{enumerate}
    \item The \textbf{HyperFlow engine} -- it is responsible for parsing the workflow representation and handling the execution order with parallelization.
    
    \item \textbf{Functions} -- they are a kind of intermediary between HyperFlow engine and cloud services. After being called by the engine, their responsibility is to create job execution requests and delegate them to the cloud via proper API. They also await the completion of the job execution, and collect metric logs from the cloud storage.
    
    \item \textbf{Handlers} -- their responsibility is to run a given job on the cloud platform. They download all files needed for job execution from cloud storage and execute them with proper arguments. Upon completion, they upload all generated data and metrics to the cloud storage.
    
    \item \textbf{Storage} -- we used cloud provider's storage to share data between tasks and collect performance metrics.
\end{enumerate}

For the purpose of our experiments, we created Fargate and Cloud Run Functions along with their Handlers, implementing the \emph{one-to-one task to container mapping} as the execution model discussed in Section~\ref{sec:implications}. We have also extended some capabilities of Lambda Function and its Handler, by adding support for execution time metrics, running JavaScript files as workflow tasks, and support for Java-based tasks. With these extensions, HyperFlow can run workflows in which tasks can be executed on Lambda, Fargate, or Cloud Run.

\subsection{Running Jobs on Serverless Containers}

Fargate and Cloud Run Handlers execute workflow jobs on Fargate Service and Cloud Run Service, respectively. They can execute various binaries in a container image in the cloud. In this paper, we created three types of container images for the handlers:

\begin{itemize}
    \item \textbf{Ubuntu Image} -- Ubuntu based image with Node.js runtime environment. It can execute custom binary files and JavaScript files.
    
    \item \textbf{Fedora Image} -- Fedora based image with Node.js runtime environment and specific libraries (e.g., Mixmod, OpenBLAS, LAPACK) configured for one of the tested applications.
    
    \item \textbf{Ubuntu Image with Java} -- Ubuntu based image with additional support for Java Archive (.jar) files.
\end{itemize}

The decision of which container image to use is made based on Function configuration, which was decorated with mapping between tasks and container images. The Function then creates a request for job execution on a given container image and sends it via proper API. Next, cloud provider provisions resources for the container and starts it. The container executes the job, uploads results files and metrics to cloud storage, and ends its work. Then, the cloud provider automatically de-provisions the container and its resources.

\section{Experiment Results}
\label{sec:experiments}

In our experiments we evaluate two major recently made available services from Amazon and Google: Fargate and Cloud Run, using four scientific workflows: Ellipsoids, Vina, KINC and Soy-KB. They were selected for evaluation since each of these workflows differs in terms of structure and resource requirements and they represent typical classes of workflows, including CPU- and data-intensive tasks. We also tested various environment setups, changing the amount of allocated memory and CPU for a given experiment. The objectives of the experiments were to:

\begin{itemize}
    \item Compare the performance of Fargate and Lambda.
    \item Compare Cloud Run and Fargate limits.
    \item Evaluate the hybrid approach that uses both FaaS and CaaS models simultaneously.
\end{itemize}

The experiments were performed on \textit{cold start} if not stated otherwise. Also both Fargate and Lambda were using a cache for container images.

\subsection{Fargate vs Lambda Comparison}
\begin{figure}[!t]
\centering
\includegraphics[width=\columnwidth]{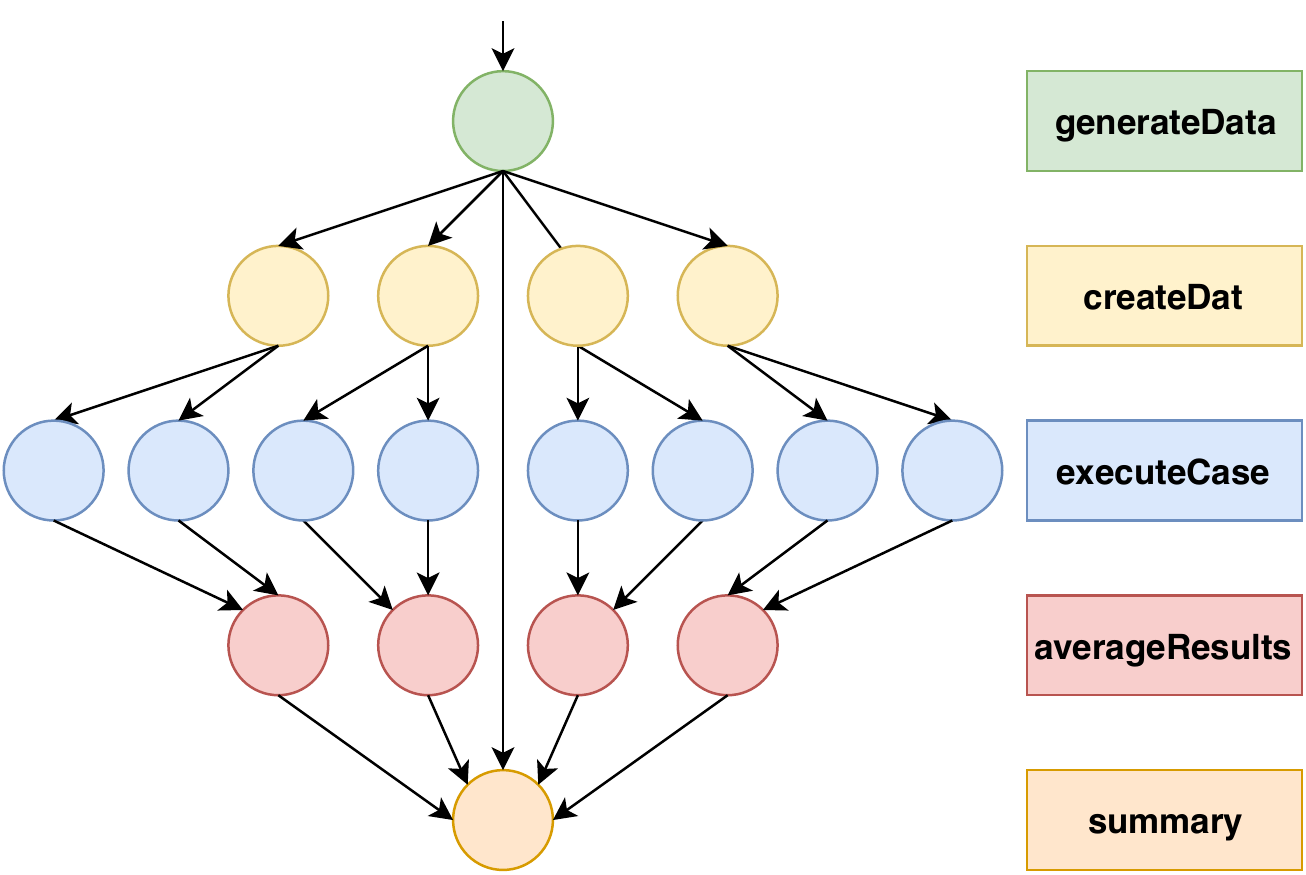}
\caption{Structure of the Ellipsoids workflow.}
\label{ellipsoids-graph}
\end{figure}

\begin{figure}[!t]
\centering
\includegraphics[scale=0.64]{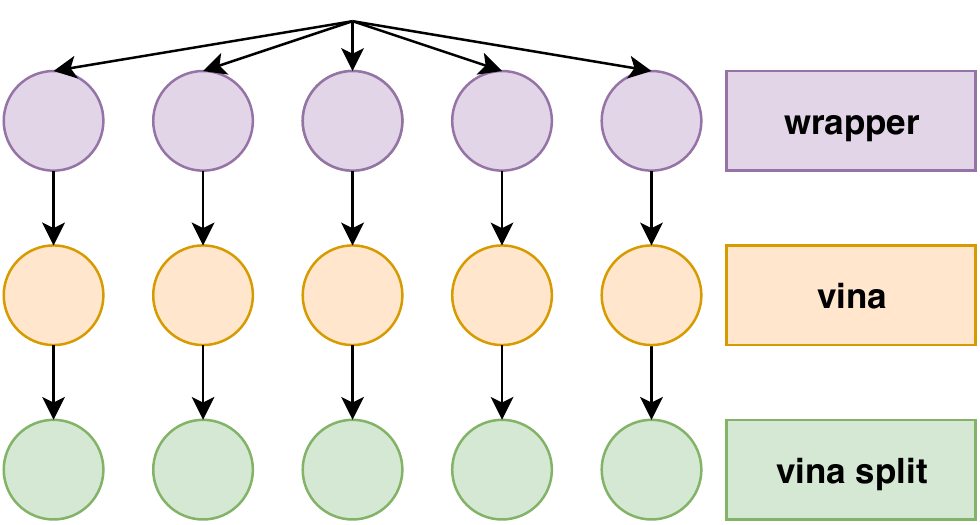}
\caption{Structure of the Vina workflow.}
\label{vina-graph}
\end{figure}

We used Ellipsoids and Vina workflows to test and compare Fargate and Lambda services. The Ellipsoids application~\cite{ellipsoids-bargiel-1, ellipsoids-bargiel-2} was created to simulate dense packing of ellipsoids in a given space, whereas AutoDock Vina~\cite{trott-vina} is an open source application for molecular docking. Both workflows consist of relatively fine-grained compute-intensive tasks (see Figs.~\ref{ellipsoids-graph} and~\ref{vina-graph}) which could be executed on both Fargate and Lambda.

\begin{figure}[!t]
\centering
\includegraphics[scale=0.52]{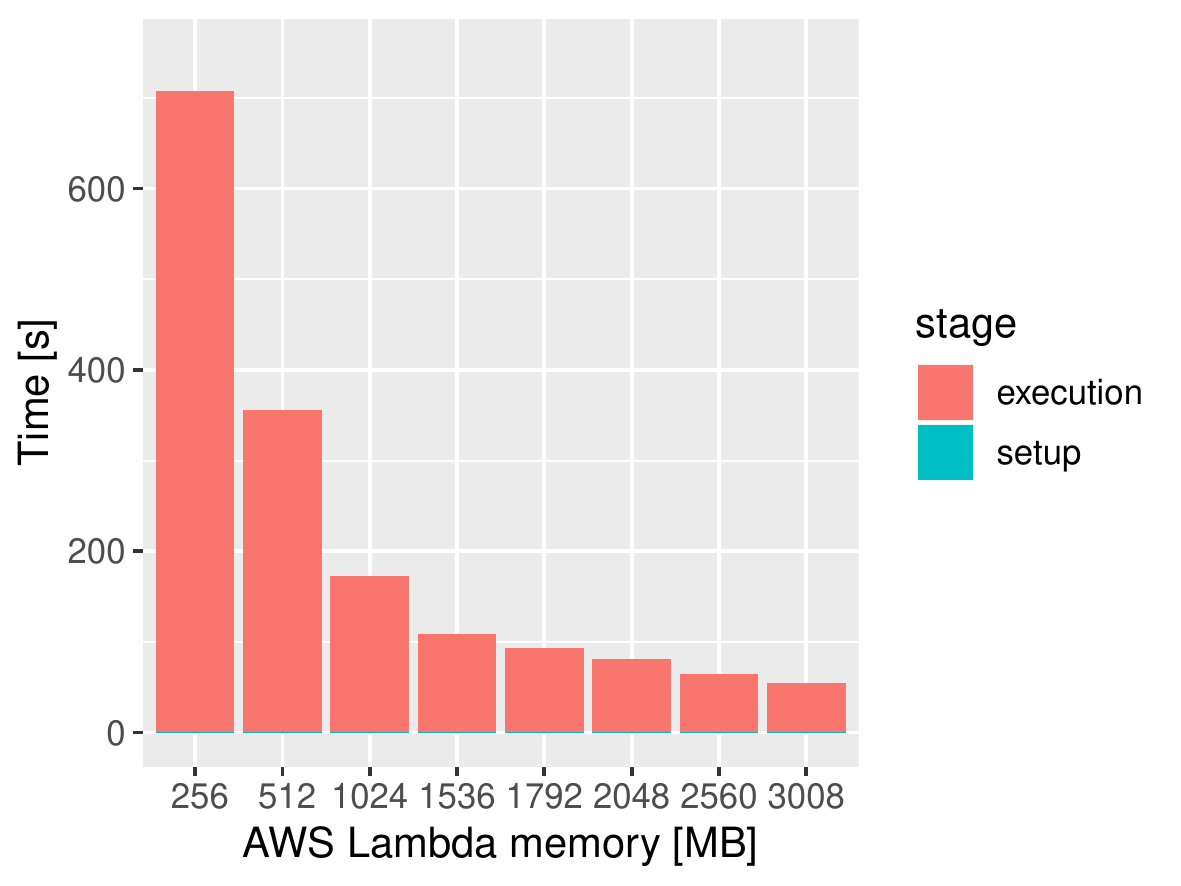}\\
\includegraphics[scale=0.52]{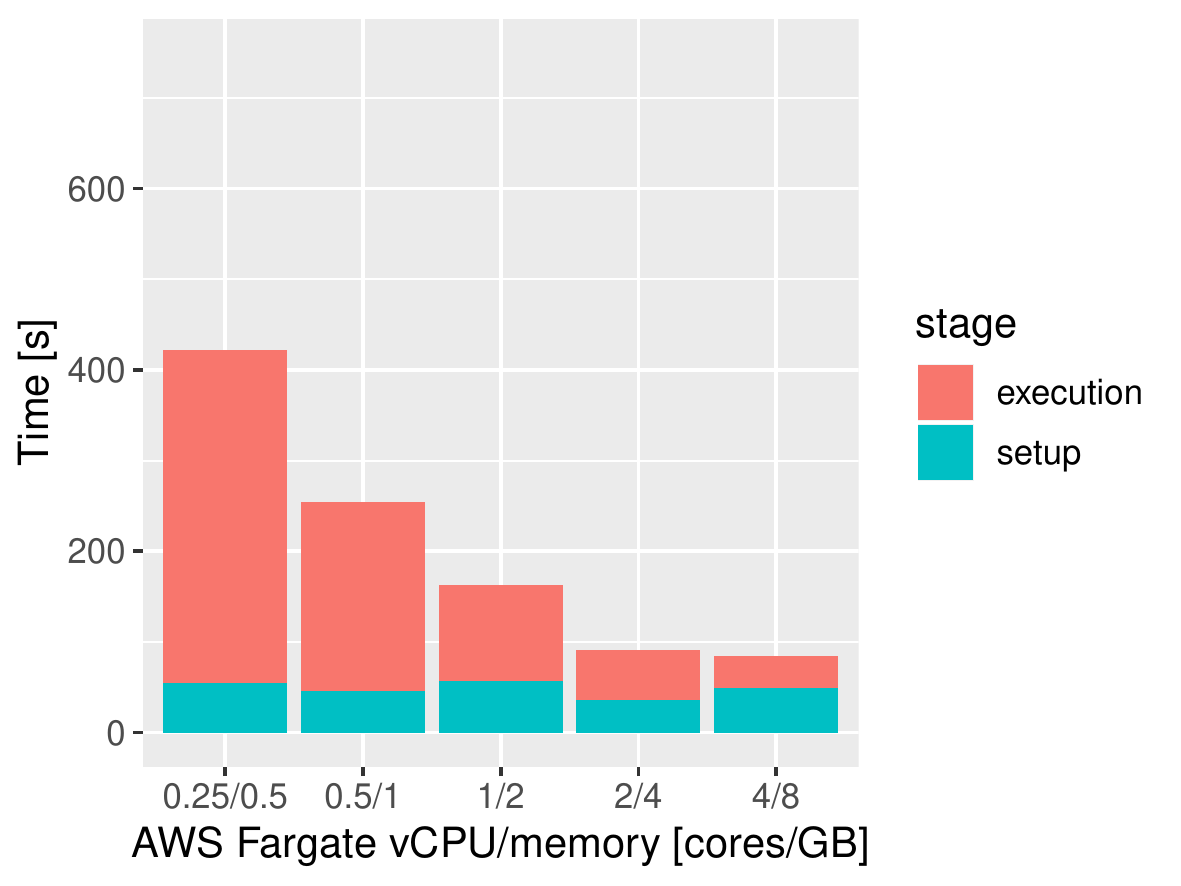}
\caption{Ellipsoids: ellipsoids-openmp task average execution time.}
\label{ellipsoids-plots}
\end{figure}

\begin{figure}[!t]
\centering
\includegraphics[scale=0.52]{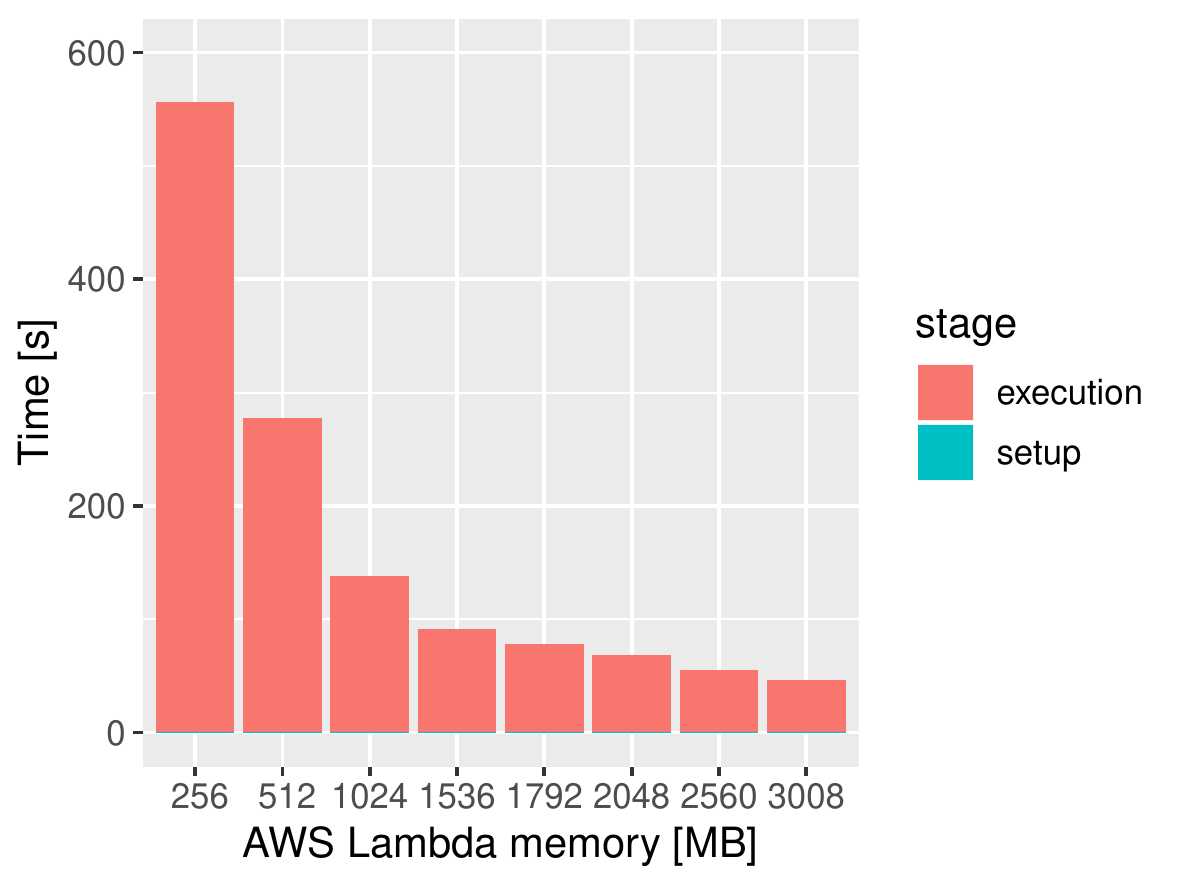}\\
\includegraphics[scale=0.52]{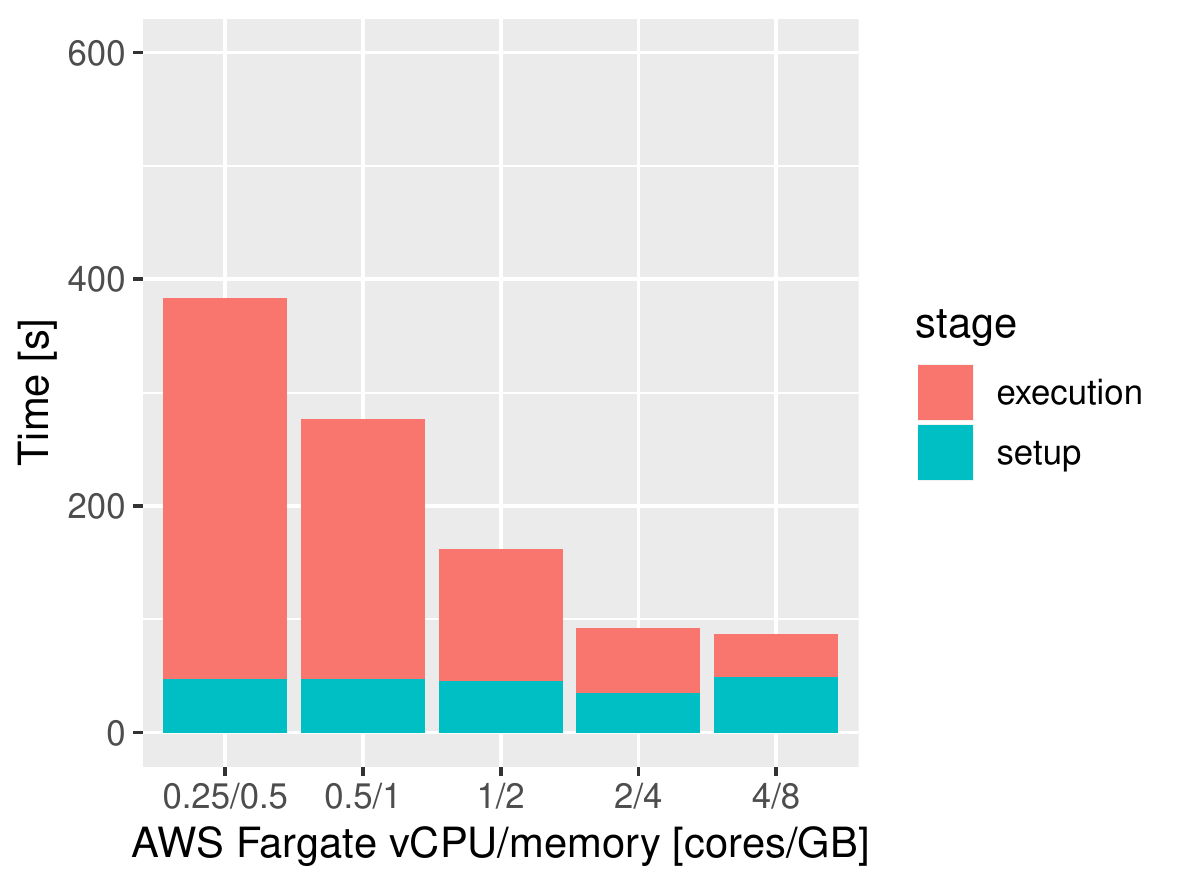}
\caption{Vina: vina task average execution time.}
\label{vina-plots}
\end{figure}

Figures~\ref{ellipsoids-plots} and~\ref{vina-plots} present the average execution time of the dominating task group (the group of tasks which takes most of the CPU time -- ellipsoids-openmp and vina tasks, respectively) in each workflow for both cloud services. For Lambda, we measure the average time for several memory configurations, while in the case of Fargate we check various vCPU/memory setups. Here and in all following Figures, \textit{setup} means the time duration between scheduling a task in HyperFlow, and starting it in the container image (handlers). The time from starting the task until its completion is \textit{execution}. When comparing runs with similar memory configurations, we can clearly see that Lambda is faster than Fargate. This is due to an overhead imposed by starting the Fargate container, which can take up to 60 seconds. This markup could be especially unfavorable for small tasks which execute in a matter of seconds. Apart from that, the Fargate container runtime also consumes some of the underlying resources like RAM or CPU time, which negatively impacts overall performance of tasks.

\begin{figure}[!t]
\centering
\includegraphics[scale=0.52]{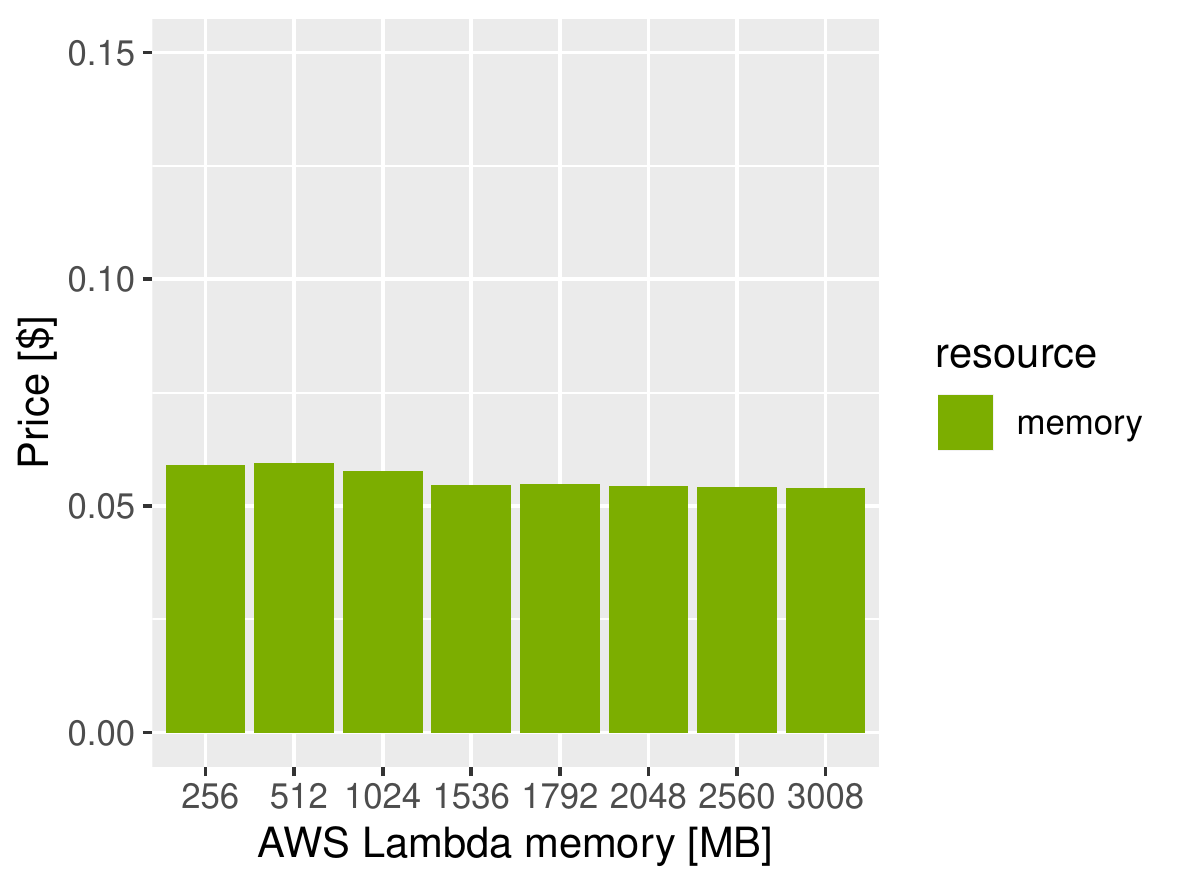}\\
\includegraphics[scale=0.52]{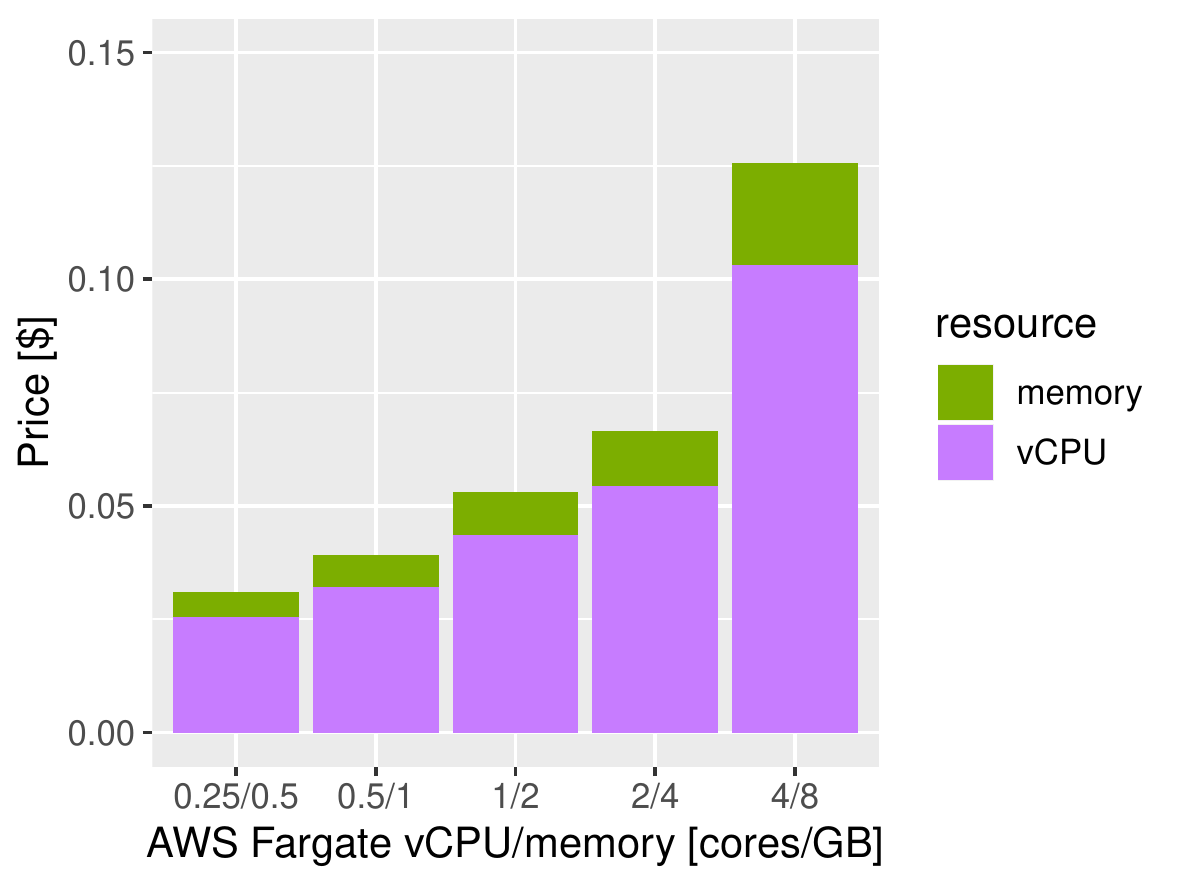}
\caption{Ellipsoids: whole workflow run cost.}
\label{ellipsoids-cost}
\end{figure}

\begin{figure}[!t]
\centering
\includegraphics[scale=0.52]{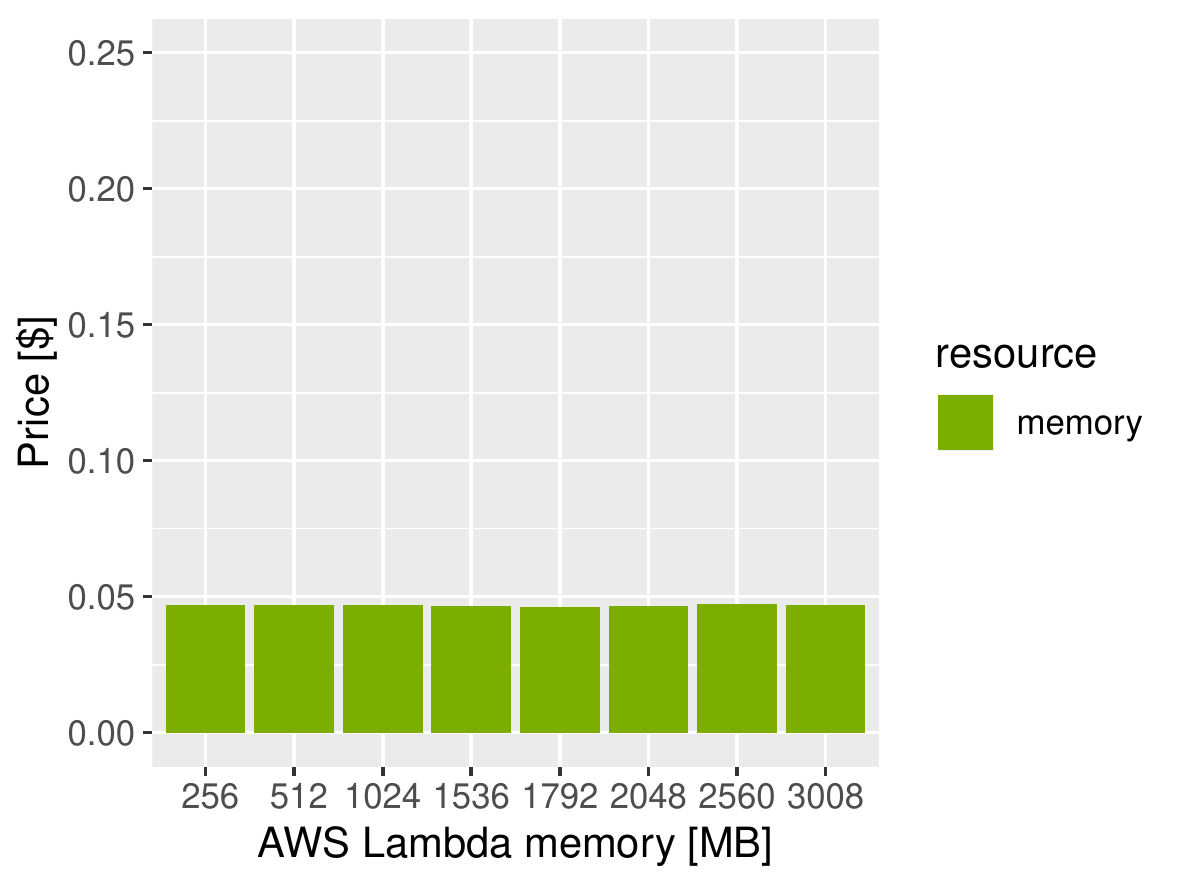}\\
\includegraphics[scale=0.52]{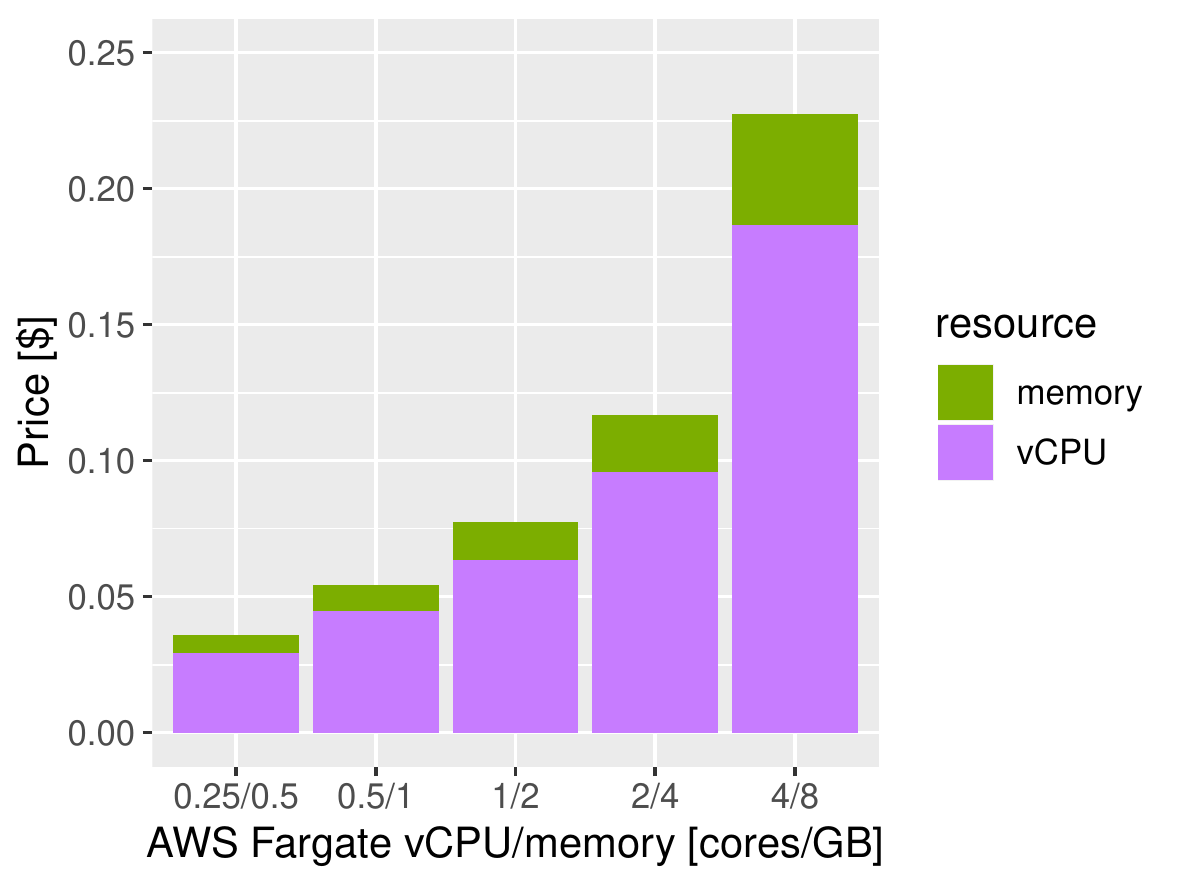}
\caption{Vina: whole workflow run cost.}
\label{vina-cost}
\end{figure}

Figures~\ref{ellipsoids-cost} and~\ref{vina-cost} show the entire cost of the workflow executions for various configurations. The cost was calculated for the EU - Ireland region, where we conducted all our experiments. Lambda is charged for each started 100ms of execution -- the price depends on the amount of allocated memory (CPU is allocated automatically by AWS); Fargate charges for each started second of execution -- pricing is based on requested vCPU and memory resources for the task. A minimum charge of 1 minute applies.
The Lambda cost for various memory configurations is similar. More expensive setups are compensated by faster execution time. It confirms if the task is CPU intensive, it is better to pick the best Lambda configuration~\cite{malawski-2017-hf-serverless,benchmarking-2}.
This is not the case for Fargate, where there is minimum memory limit for each vCPU value. Considering our tasks are CPU sensitive, this means we have to pay for the memory even though it does not improve performance. In addition, adding more vCPUs after some threshold does not significantly improve performance. We can draw a conclusion that Fargate is not best suited for small-grained, low memory tasks. In this case, Lambda will perform better. On the other hand, we must recall that Fargate does not have constraints like the time limit of the execution task and we can allocate many more resources such as memory or disk space.
\subsection{Fargate vs. Cloud Run Limits}

\begin{figure}[!t]
\centering
\includegraphics[width=\columnwidth]{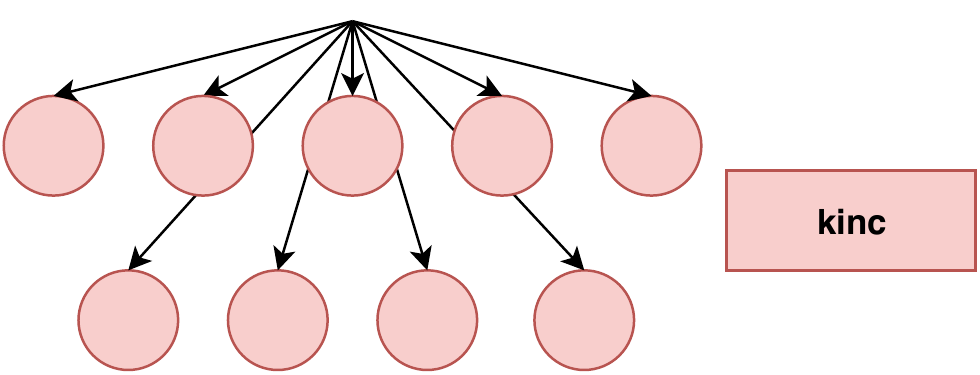}
\caption{Structure of the KINC workflow.}
\label{kinc-graph}
\end{figure}

In our next experiment, we used the OSG-KINC workflow to determine two important serverless features -- elasticity and scalability of Fargate and Cloud Run Services.
OSG-KINC workflow~\cite{osg-kinc-paper}, is a  software to build a similarity matrix representing correlation analysis of all pairwise comparisons of genes/transcripts in a gene expression matrix. It is a data-intensive workflow that is configured to run KINC -- Knowledge Independent Network Construction using Pegasus on the Open Science Grid. Its main characteristic is that it has only one type of task, but potentially many of them, reaching thousands (see Fig.~\ref{kinc-graph}). Thus it is a good candidate to measure elasticity of cloud services. Moreover, the size of input data (genetic sequences in the size of gigabytes) prohibits running them on AWS Lambda. We transformed the workflow format so that it could be executed by HyperFlow and packaged it into containers.

\begin{figure}[!t]
\centering
\includegraphics[width=\columnwidth]{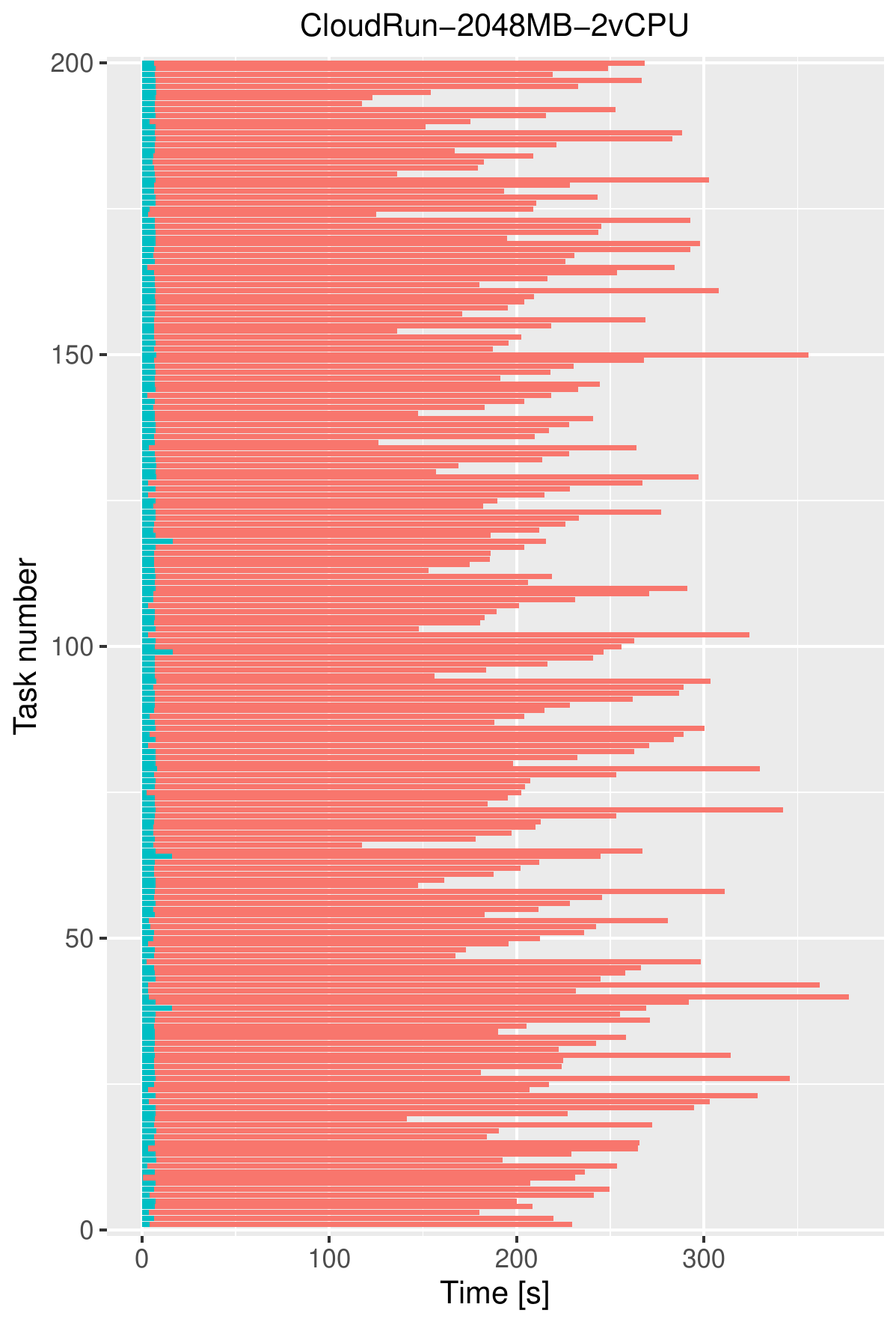}
\caption{OSG-KINC on Cloud Run: chart.}
\label{kinc-gcloud}
\end{figure}

\begin{figure}[!t]
\centering
\includegraphics[width=\columnwidth]{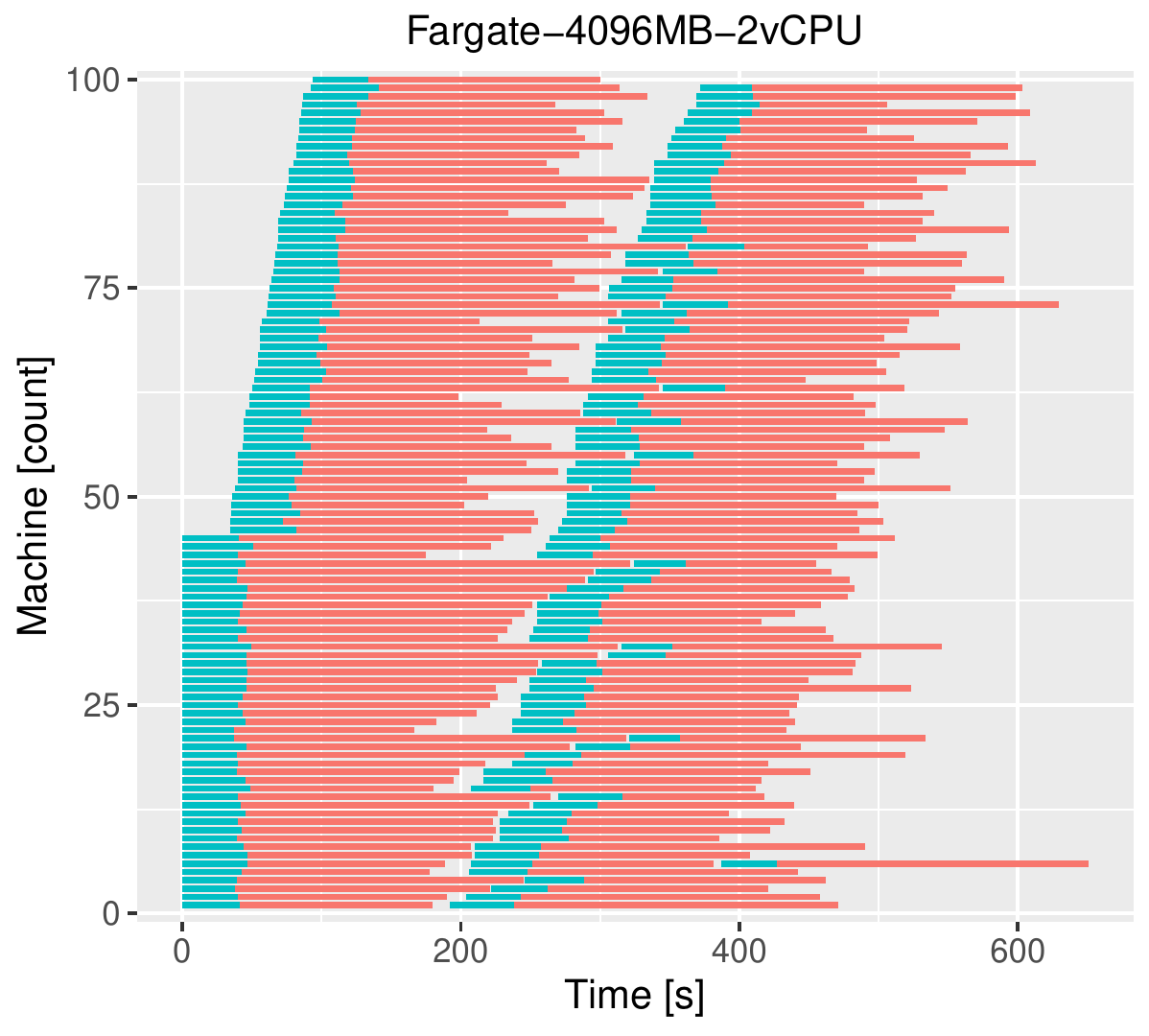}
\caption{OSG-KINC on Fargate: flattened Gantt chart.}
\label{kinc-fargate}
\end{figure}

\begin{figure}[!t]
\centering
\includegraphics[width=\columnwidth]{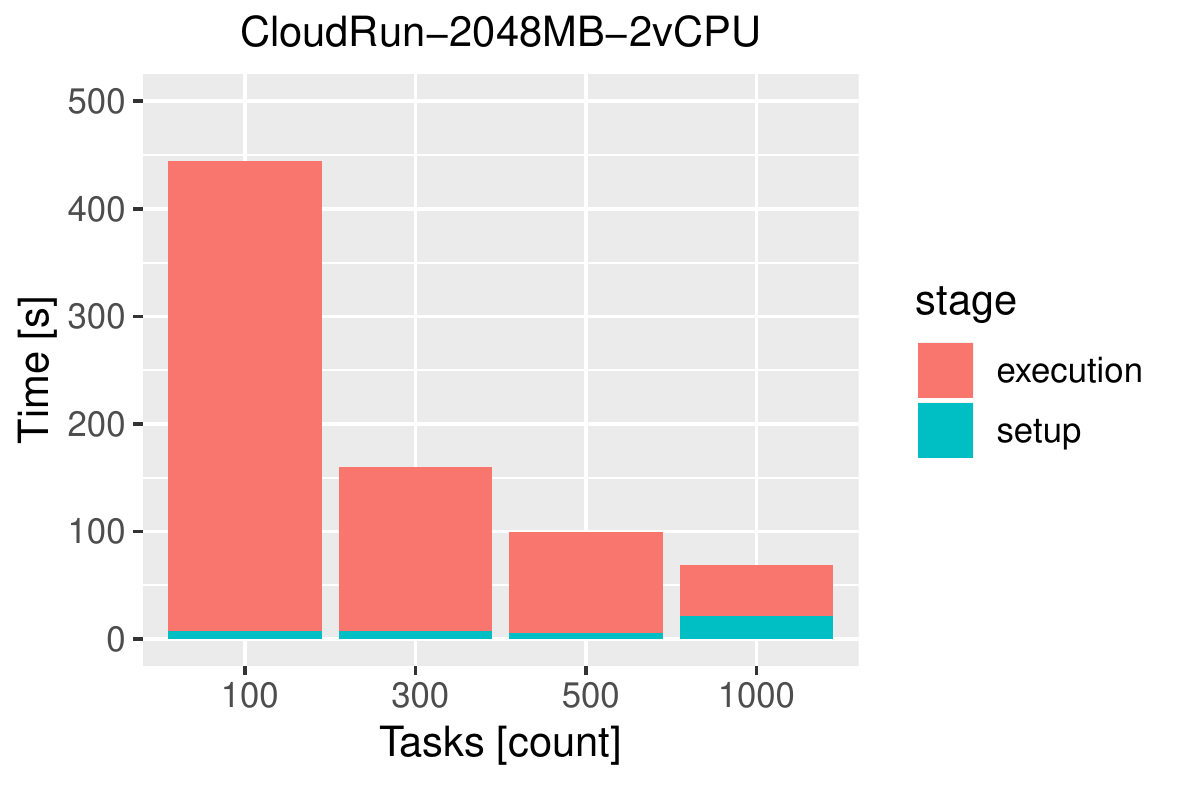}
\caption{OSG-KINC: kinc task average execution time.}
\label{kinc-plot}
\end{figure}

\begin{figure}[!t]
\centering
\includegraphics[width=0.8\columnwidth]{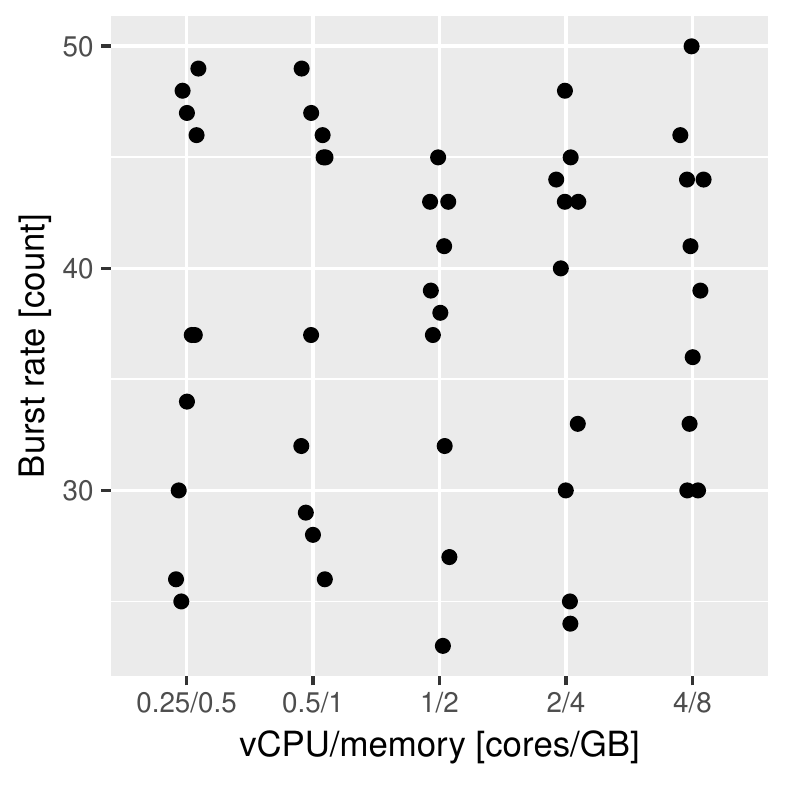}
\caption{OSG-KINC: AWS Fargate burst rate.}
\label{kinc-fargate-burst}
\end{figure}

\begin{figure}[!t]
\centering
\includegraphics[width=\columnwidth]{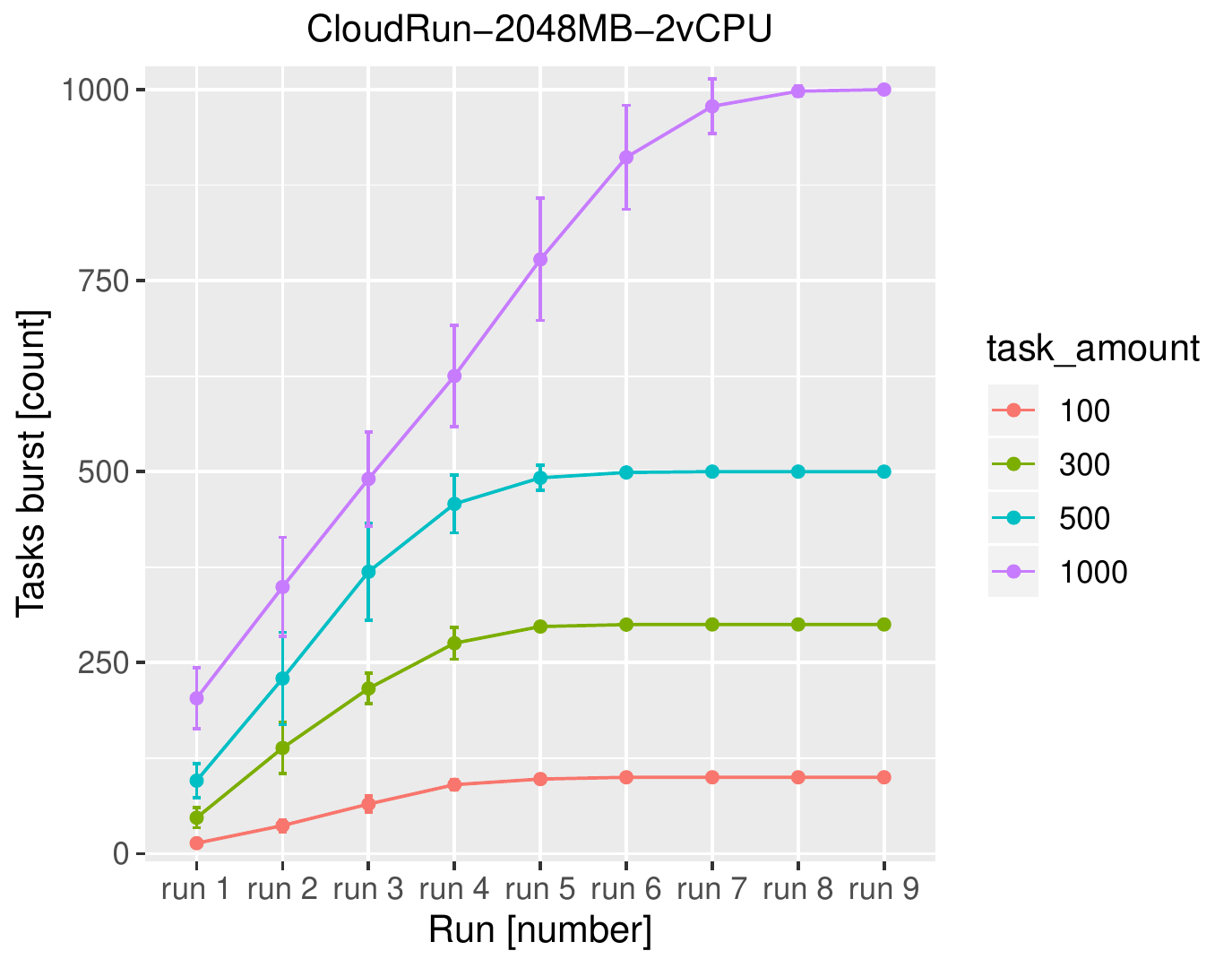}
\caption{OSG-KINC: Cloud Run burst rate.}
\label{kinc-cloudrun-burst}
\end{figure}

We run the KINC workflows of 4 sizes (100, 300, 500 and 1000 tasks), repeating each execution 10 times. Figures~\ref{kinc-gcloud} and~\ref{kinc-fargate} show examples of execution charts of workflows with 200 KINC tasks for both services. We can note that execution time on average is slightly longer on Cloud Run. This is because Fargate had faster physical cores ($\sim$2.9~GHz) than Cloud Run ($\sim$2.35~GHz) in our experiments.

With Cloud Run execution (Figure~\ref{kinc-gcloud}) there were not any significant issues or delays. Some individual tasks took more time for setup (blue line) due to waiting for container spawning. Figure~\ref{kinc-plot} presents the average execution time with distinction between setup and execution time for various tasks amount. We can notice that only for 1000 tasks the setup time is longer on average than for smaller amount of tasks, it takes longer for Cloud Run scale-up the infrastructure for so many tasks.

Figure~\ref{kinc-fargate} shows an example flattened Gantt chart for the Fargate execution. Although all tasks were spawned by the HyperFlow all at once, Fargate did not start them straightaway. There are two explanations for this:

\begin{itemize}
    \item The first issue is related with Fargate limits. We cannot start more than 100 tasks at a given moment. After firing the first 100 tasks, we need to wait for some tasks to complete before going further.

    \item The second observation, which is more interesting, is related to the Fargate burst rate. If we take a closer look at the chart, we can see that the first 100 tasks did not start at the same time. Although the task limit is set to 100, we could not burst running tasks from 0 to 100 at once. After some point, we started obtaining \textit{ThrottlingException: Rate Exceeded} errors when submitting tasks. After this error, we had to wait for some time and retry it in a repeatable manner. At the moment, AWS does not specify any details or limitations related to this problem. We have then decided to check whether this issue is somehow related to the chosen configuration. 
\end{itemize}

Figure~\ref{kinc-fargate-burst} presents the measured burst rate from 10 executions on each configuration as a scatter plot. Note that we jittered the points because some values were overlapping. On average, the burst rate for all setups is about 38 tasks. It demonstrates that the problem is not related with the amount of resources but rather with some limitation of AWS API. This is unfortunate because from a serverless service like Fargate we would expect a high level of scalability and elasticity.

We performed similar measurement for Cloud Run shown in Figure ~\ref{kinc-cloudrun-burst}. However, this time we observed that with each consecutive run we are able to handle more and more tasks all at once (beginning with cold start, later without it). This is surely affected by different design choices behind Cloud Run where the containers are more lightweight in comparison with Fargate. In addition, Cloud Run containers are expected to expose server endpoint which handles requests (tasks) and be reusable after execution (statelessness). It makes them more suitable as serverless solution in some situations, but comes at a price of smaller resource pool and maximum execution time limit.

\subsection{Hybrid Approach -- Fargate and Lambda}
\label{sec:hybrid}
In our last experiment, we evaluated the Soybean Knowledge Base~\cite{soykb-paper}, an application used for in depth analysis of the genotypic data of soybeans. It sequences large sets of germplasm in crops in order to detect genome-scale genetic variations. The results of this analysis can be used to better understand and study various traits for the improvement of crops by design. 

\begin{figure}[H]
\centering
\includegraphics[scale=0.64]{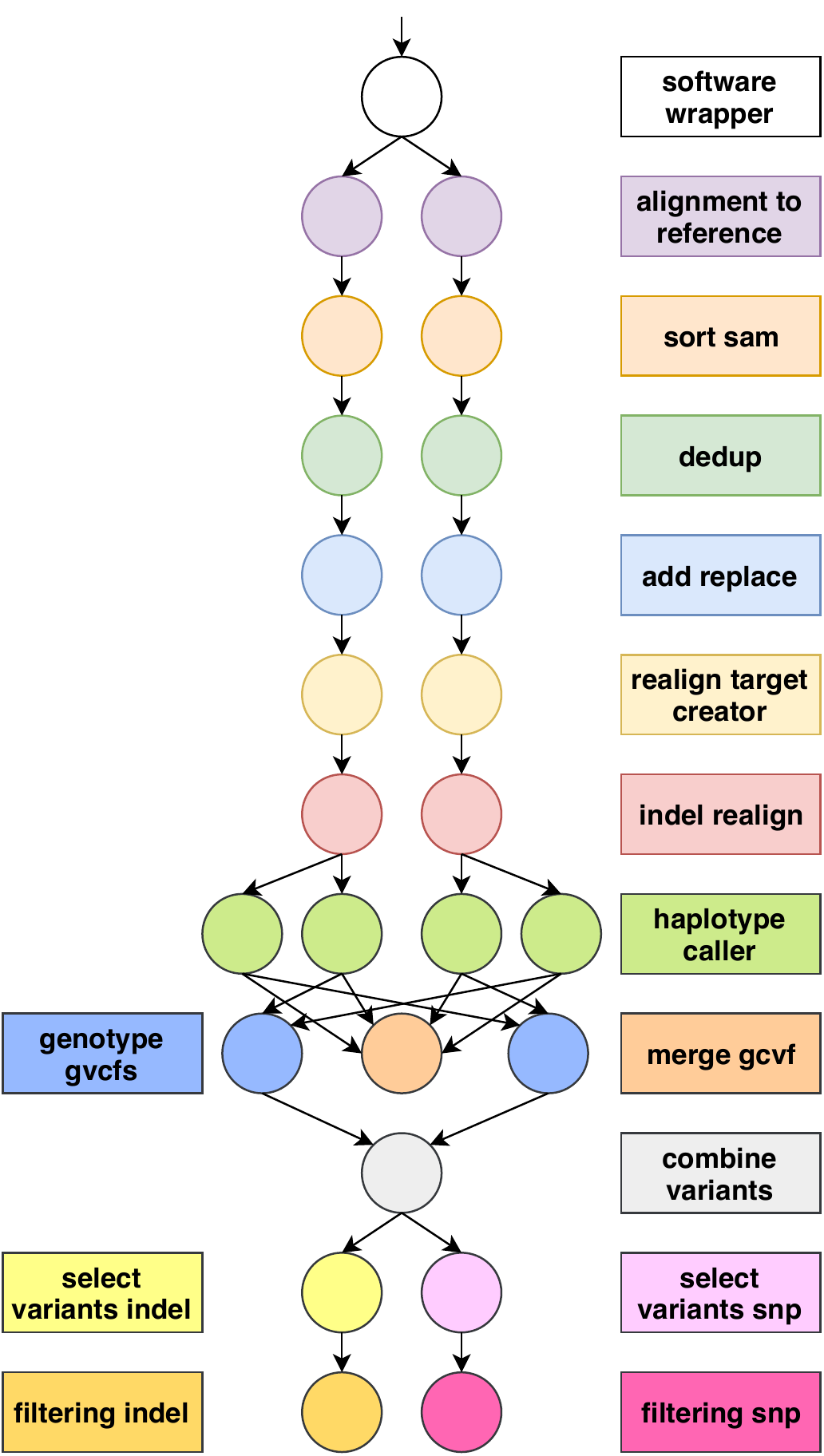}
\caption{Structure of the SoyKB workflow.}
\label{soykb-graph}
\end{figure}

The main  distinction of SoyKB workflow from others is its complexity. It has many stages and the number of tasks spans from tens to thousands depending on the given case and configuration (see Fig.~\ref{soykb-graph}). Also, the task execution times are quite different, depending on the stage and given input. We wanted to find out how well Fargate can cope with such big workflows and determine whether a hybrid approach of using both CaaS and FaaS services simultaneously is feasible.

\begin{figure}[H]
\centering
\includegraphics[width=\columnwidth]{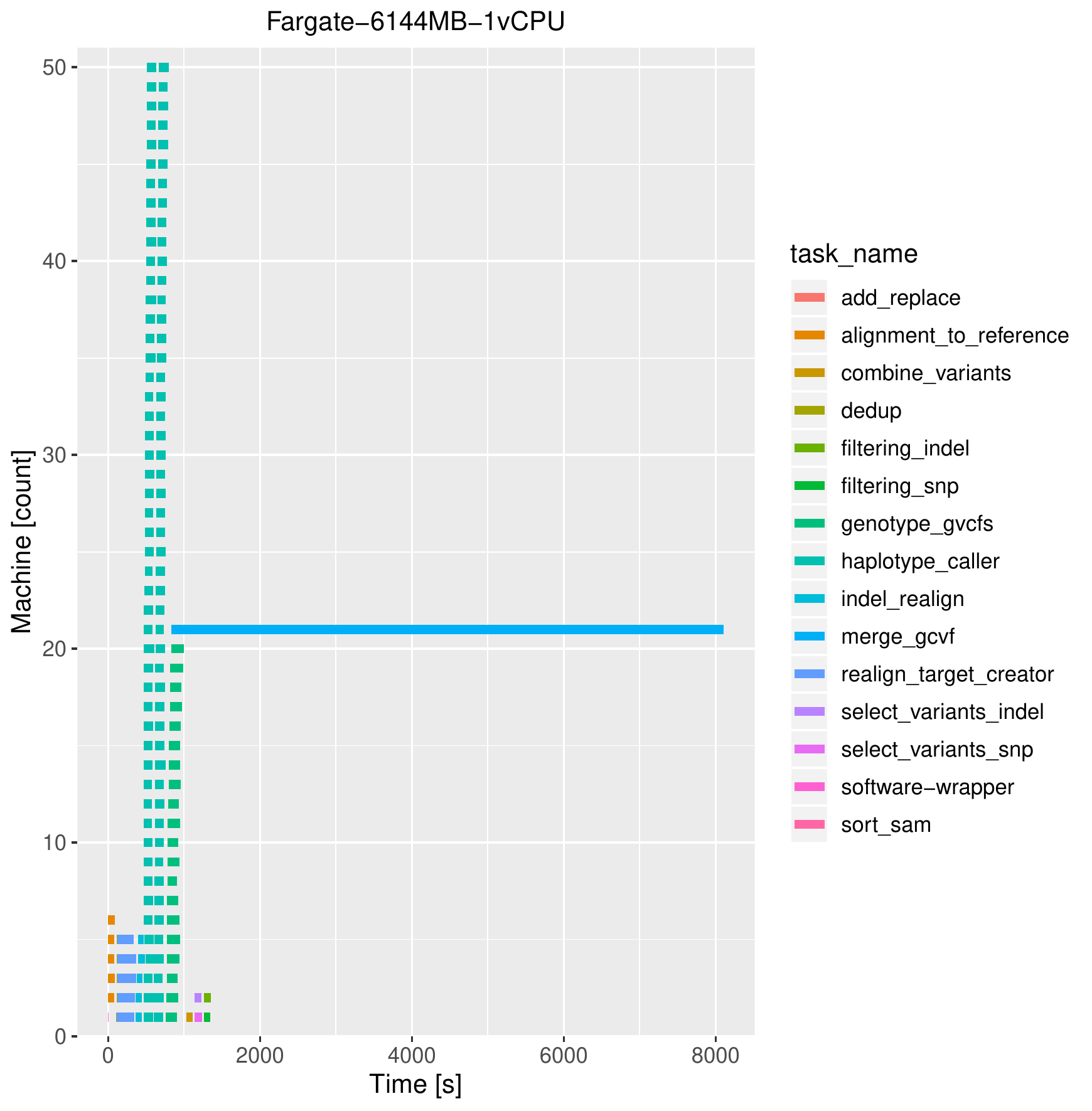}
\caption{SoyKB: flattened Gantt chart.}
\label{soykb-gantt}
\end{figure}

Figure~\ref{soykb-gantt} shows the results of the hybrid execution. With HyperFlow engine, we managed to use Lambda and Fargate concurrently. Lambda runs small-grained tasks; \textit{software-wrapper}, \textit{sort{\_}sam}, \textit{dedup}, and \textit{add{\_}replace}. The remaining tasks are executed on Fargate. Thanks to this approach, we obtained results from small-grained tasks in a matter of seconds. On the other hand, Fargate handled more demanding tasks which exceeded Lambda limits. The same could be achieved with our own cluster of running machines (e.g., Amazon EC2), but it would require additional work to setup the infrastructure.

This experiment demonstrated that hybrid approach is feasible and it can be quite effective. Both services complement each other quite well. We have also demonstrated that Fargate can be utilized for workflows with more variety in task requirements, especially those workflows which include long running tasks. Of course final decision on what to use also depends on services limits. For example, Cloud Run at the moment does not allow enough memory for a container to handle large tasks from this workflow.
\section{Conclusions and Future Work}
\label{sec:conclusion}

In this paper, we introduced the serverless containers on the example of AWS Fargate and Google Cloud Run, as a new viable option for running scientific workflows. We have discussed the execution models for workflow tasks on this type of infrastructure, and compared it to traditional models as well as to cloud functions on the example of AWS Lambda. Then, we implemented the \emph{one-to-one task to container mapping}, as the execution model which takes most advantages of the pure serverless infrastructure. Thanks to this approach, the user (or the workflow management system) does not need to make complex auto-scaling decisions and worry about over- and under-provisioning. Up to our knowledge, this is the first report on running container-based scientific workflows in a fully serverless model.

Our experiments with the implementation based on HyperFlow engine, and four real-world  workflows using the two leading cloud providers (Amazon and Google), confirmed the applicability of serverless containers services for scientific workflows and specific advantages of such platforms. They include the possibility to deploy own container images with scientific software, no execution time limits, and disk and memory quotas sufficient for our applications, including a data-intensive genomics workflow (SoyKB). We confirmed the current scalability limits for Fargate and Cloud Run and observed the expected speedup proportional to the allocated vCPU share. On the other hand, we observed that currently Fargate introduces a container setup overhead of about 60 seconds, and that the API has a throttling limit, which prevents large bursts of concurrent tasks (currently between 30 and 50 requests). On Cloud Run these issues are less severe, but there are other constraints in the form of execution time limit or smaller resource pool.

\begin{table*}[!t]
\begin{center}
\caption{Comparison of evaluated cloud service models.}
\begin{tabular}{ >{\centering\arraybackslash}m{4.2cm} >{\centering\arraybackslash}m{4.2cm} >{\centering\arraybackslash}m{4.2cm}} 
 \toprule
 \textbf{CaaS} & \textbf{FaaS} & \textbf{Hybrid} \\ 
 \midrule
  Serverless &  Serverless & Serverless \\
 \midrule
  Runs containers &  Runs functions & Runs containers and functions \\
 \midrule
 Scalable &  Well-scalable & Well-scalable \\
 \midrule
 Moderate quotas limits &  Major quotas limits & Moderate quotas limits \\
  \midrule
 Minor execution time limits &  Major execution time limits & Minor execution time limits \\
 \bottomrule
\end{tabular}
\label{cloud-services-comparison}
\end{center}
\end{table*}

Regarding the price and performance comparison of serverless containers and Lambda, it turns out that Lambda is more cost-efficient for compute-intensive tasks, provided they fit into the resource limits of Lambda. In general, using serverless containers is not recommended for tasks shorter than several minutes, since the setup overhead becomes prohibitive for smaller tasks. A general conclusion is thus to use a hybrid approach, because it takes the best properties from both CaaS and FaaS models (see Table~\ref{cloud-services-comparison}). As it was shown in Section~\ref{sec:hybrid}: use Lambda for all the tasks that fit into resource limits, and the others that are either too long, too heavy in terms of disk/memory or require more complicated software setup with a custom image, should be run on Fargate or Cloud Run. 

We claim that our work is general for other serverless computing platforms, as the services from other providers such as Azure Serverless Containers have similar properties to Fargate and Cloud Run. This still needs to be properly evaluated, and is the subject of our future work. There are also other new platforms to manage modern serverless workloads such as Knative\footnote{https://knative.dev/}, Kubeless\footnote{https://kubeless.io/}, or OpenWhisk\footnote{https://openwhisk.apache.org/} which could be worth researching. In our evaluation, all the tasks within a single workflow were run on the containers of the same size, whereas future research focuses on development of scheduling algorithms for scientific workflows based on the performance evaluation we conducted here. Moreover, our prototype implementation based on HyperFlow needs to be extended for production use and ported to other workflow management systems, including Pegasus.

\subsubsection*{Acknowledgments}
This work was supported by the National Science Centre, Poland, grant 2016/21/B/ST6/01497.

\bibliographystyle{elsarticle-num}
\bibliography{references}

\par\noindent 
\parbox[t]{\linewidth}{
\noindent\parpic{\includegraphics[height=1.4in,width=1in,clip,keepaspectratio]{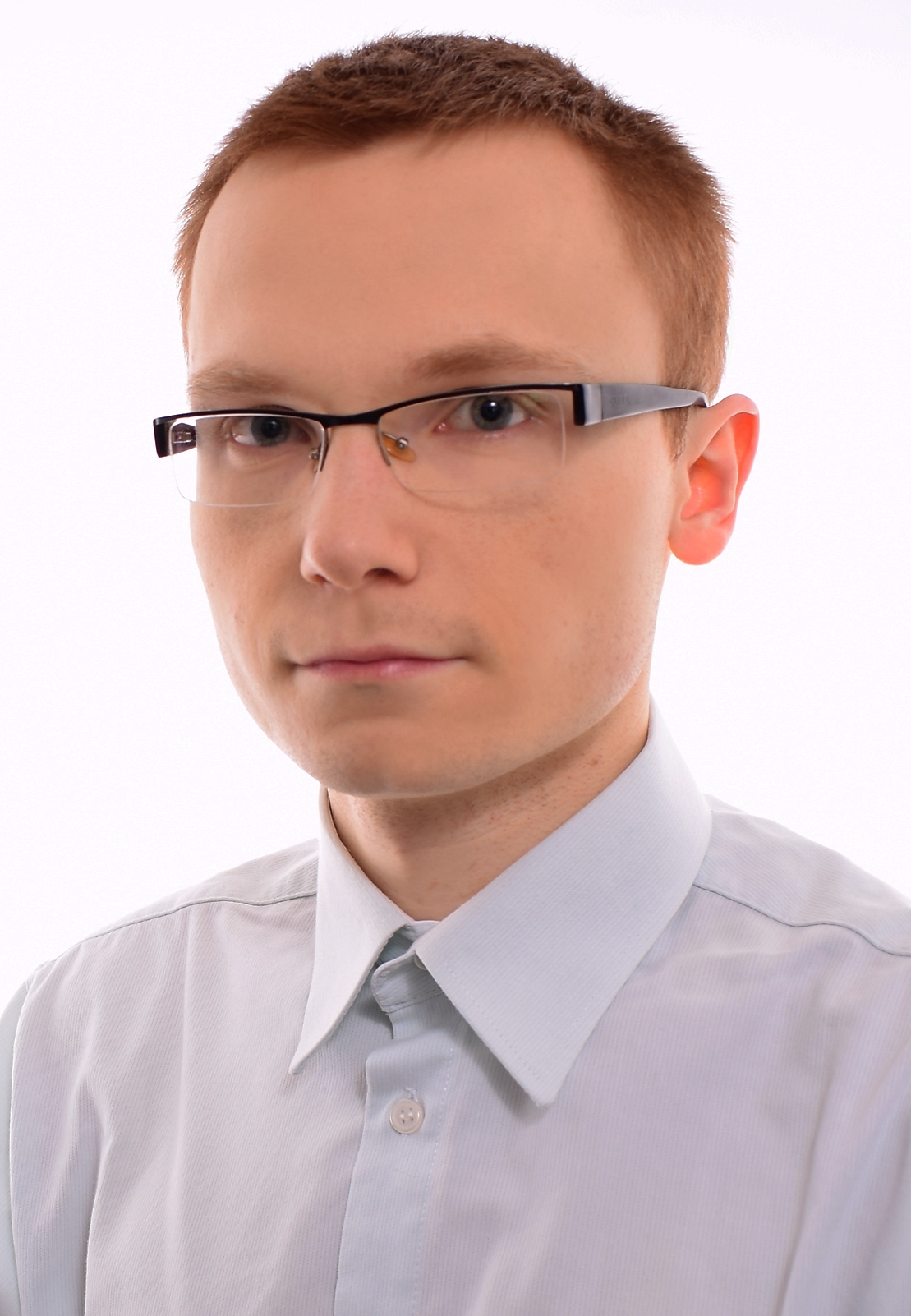}}
\noindent {\bf Krzysztof Burkat}\
holds a M.Sc. in Computer Science from the AGH University of Science and Technology in  Krak{\'o}w, Poland and works as a Software Engineer for UBS Investment Bank in a big data project. He also conducts research in serverless computing as a part of scientific grant from National  Science  Centre, Poland. His  main  scientific interests include cloud computing, big data, software engineering and distributed systems. He previously worked at Sabre Airline Solutions.}
\vspace{2\baselineskip}

\par\noindent 
\parbox[t]{\linewidth}{
\noindent\parpic{\includegraphics[height=1.4in,width=1in,clip,keepaspectratio]{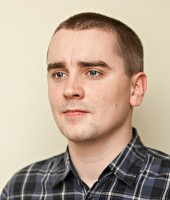}}
\noindent {\bf Maciej Pawlik}\
holds a M.Sc. in Applied Computer Science from University of Science and Technology. He is a PhD student at the Department of Computer Science AGH, where he's involved in European and national research projects. His main areas of interest are: novel cloud infrastructures, performance measurement and modeling, workflow engines and scheduling of distributed applications.}
\vspace{2\baselineskip}

\par\noindent 
\parbox[t]{\linewidth}{
\noindent\parpic{\includegraphics[height=1.4in,width=1in,clip,keepaspectratio]{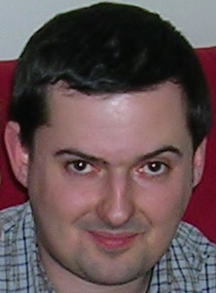}}
\noindent {\bf Bartosz Bali\'s}\
holds a PhD in Computer Science from the AGH University of Science and Technology. He is an associate professor at the Department
  of Computer Science AGH. He is a co-author
   of around 100 international scientific publications. His research interests include scientific workflows, e-Science, cloud computing, and distributed computing.}
\vspace{2\baselineskip}

\par\noindent 
\parbox[t]{\linewidth}{
\noindent\parpic{\includegraphics[height=1.4in,width=1in,clip,keepaspectratio]{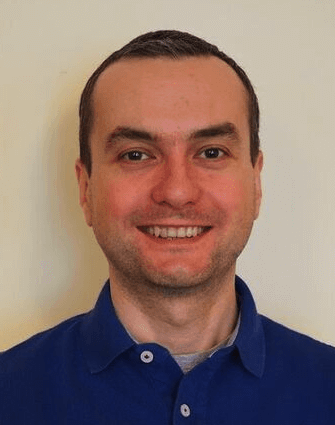}}
\noindent {\bf Maciej Malawski}\
holds a PhD in computer science from AGH, in 2011-2012 he was a postdoc at the University of Notre Dame, USA. Currently he is an associate professor at the Department of Computer Science AGH. His scientific interests include parallel computing, grid and cloud technologies, distributed systems, resource management and scientific applications.}
\vspace{2\baselineskip}

\par\noindent 
\parbox[t]{\linewidth}{
\noindent\parpic{\includegraphics[height=1.4in,width=1in,clip,keepaspectratio]{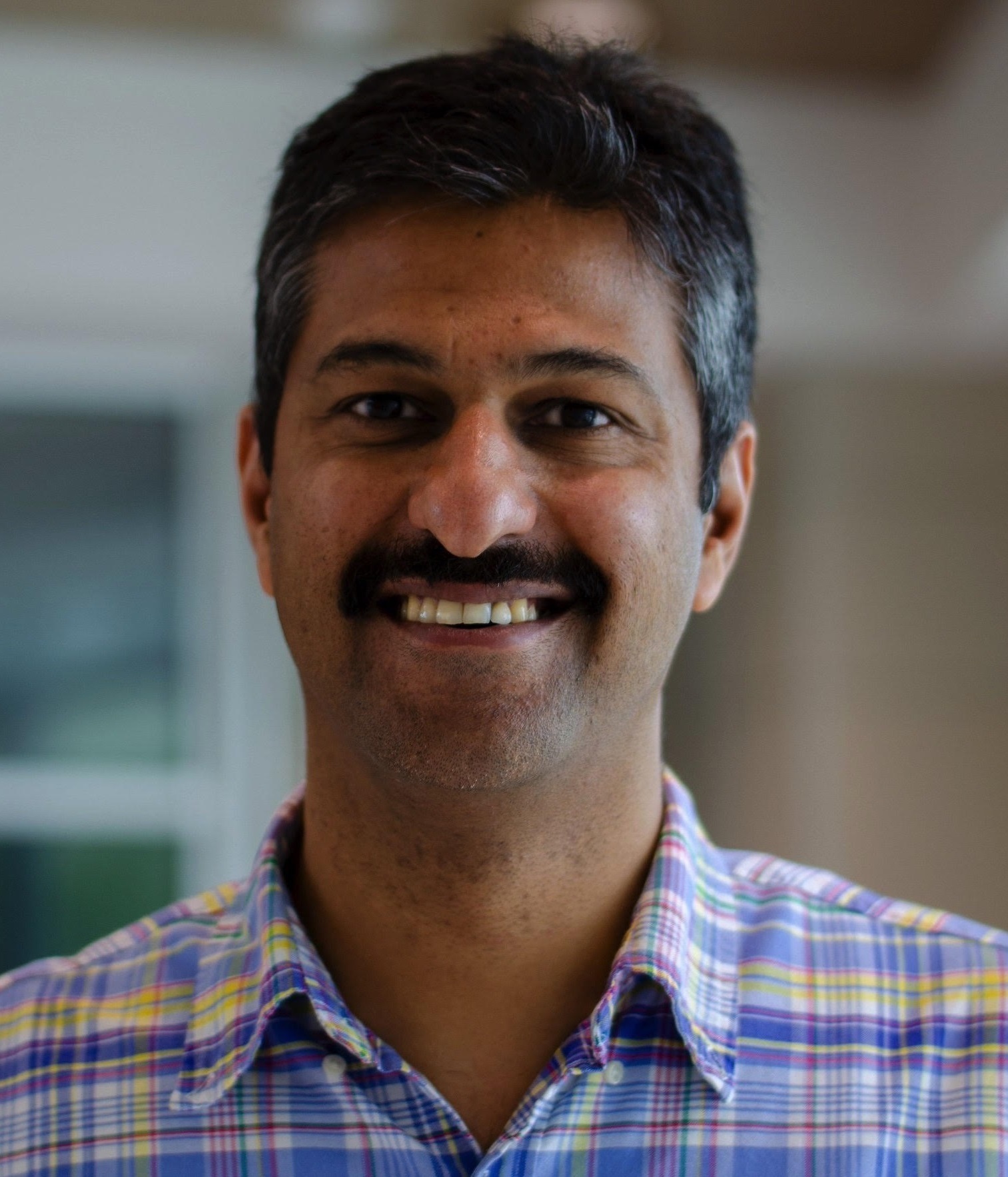}}
\noindent {\bf Karan Vahi}\
is a Senior Computer Scientist in the Science Automation Technologies group at ISI. He is the lead developer of Pegasus and is in charge of the core development. Karan has been working on Pegasus since 2002. His research interests include distributed and high-performance computing. Mats has a MS in computer science from University of Southern California, Los Angeles.}
\vspace{2\baselineskip}

\par\noindent 
\parbox[t]{\linewidth}{
\noindent\parpic{\includegraphics[height=1.4in,width=1in,clip,keepaspectratio]{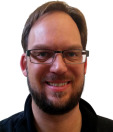}}
\noindent {\bf Mats Rynge}\
is a computer scientist at the University of Southern California's Information Sciences Institute. His research interests include distributed and high-performance computing. He received the B.S. degree in computer science from the University of California, Los Angeles.}
\vspace{2\baselineskip}

\par\noindent 
\parbox[t]{\linewidth}{
\noindent\parpic{\includegraphics[height=1.4in,width=1in,clip,keepaspectratio]{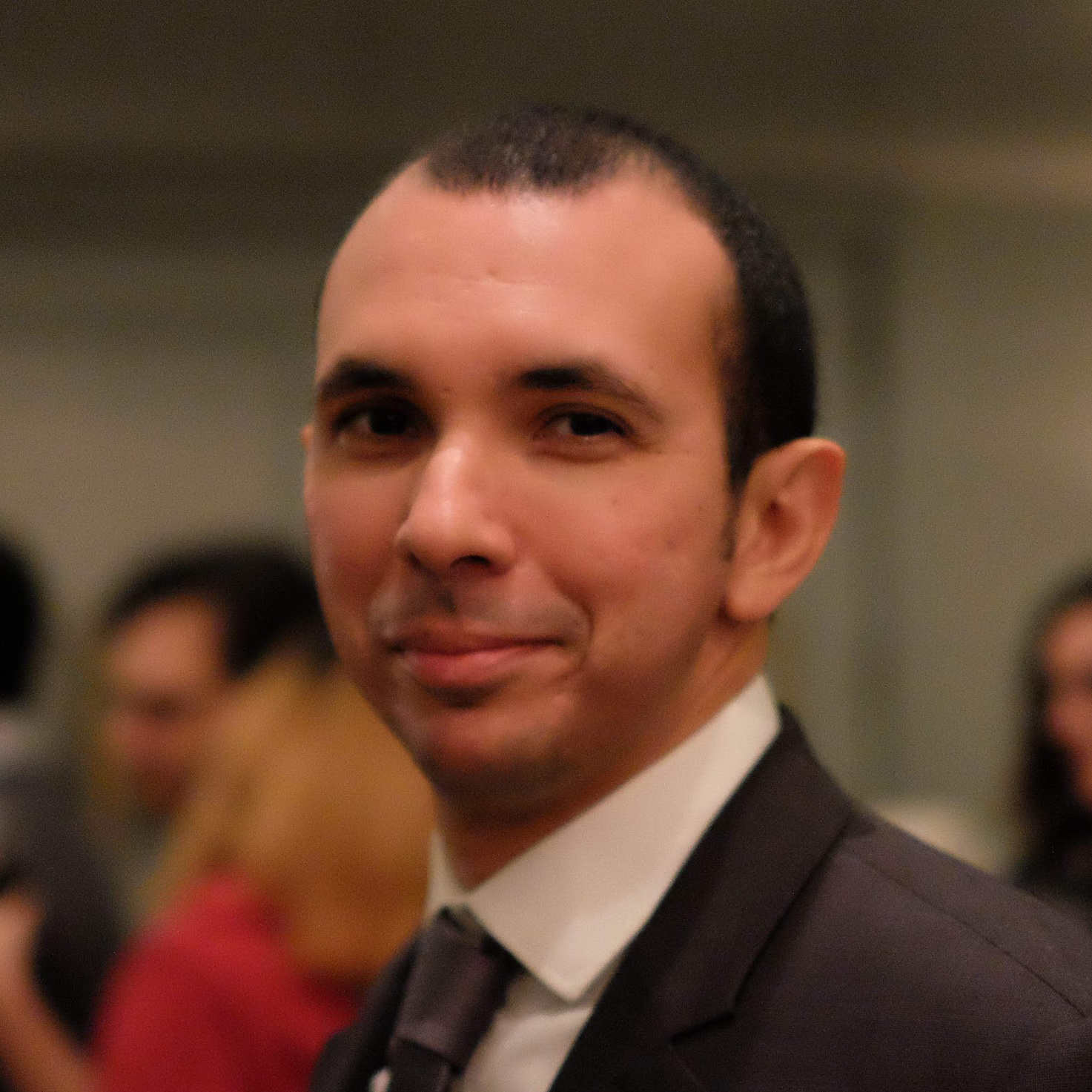}}
\noindent {\bf Rafael Ferreira da Silva}\
is a Research Assistant Professor in the Department of Computer Science at University of Southern California, and a Research Lead in the Science Automation Technologies group at the USC Information Sciences Institute. His research focuses on the Modeling and Simulation of Parallel and Distributed Computing Systems, and efficient execution of Scientific Workflows on heterogeneous distributed systems (e.g., clouds, grids, and supercomputers). He received his PhD in Computer Science from INSA-Lyon, France.}
\vspace{2\baselineskip}

\par\noindent 
\parbox[t]{\linewidth}{
\noindent\parpic{\includegraphics[height=1.4in,width=1in,clip,keepaspectratio]{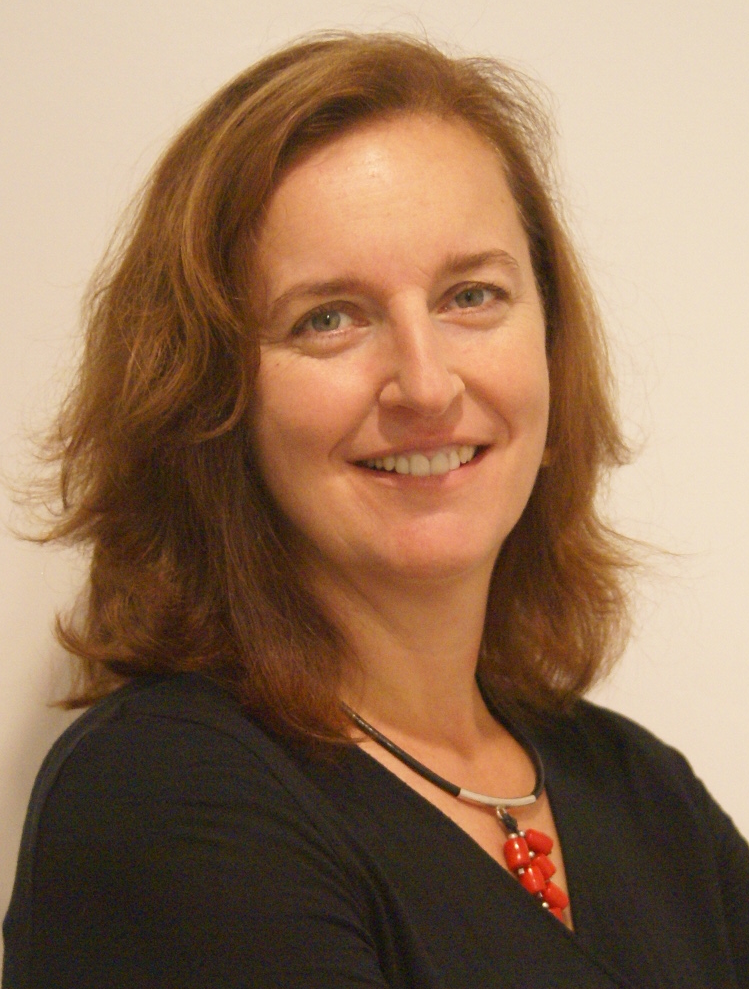}}
\noindent {\bf Ewa Deelman}\
received the PhD degree in computer science from Rensselaer Polytechnic Institute. She is a Research Director of the Science Automation Technologies group in the Advanced Systems Division at ISI and a Research Professor in the Dept of Computer Science at University of Southern California. Her research focuses on distributed computing, in particular regarding how
to best support complex scientific applications on a
variety of computational environments, including
campus clusters, grids, and clouds.}
\vspace{2\baselineskip}

\end{document}